\documentclass[11pt,a4paper, amsmath, fleqn]{article}
\usepackage{jcappub}
\usepackage[english]{babel}
\usepackage{graphicx,amsmath,amsfonts,amssymb,slashed,upgreek,color,caption, tensor, epsfig}
\usepackage{array}
\usepackage{subfigure}
\usepackage[numbers,sort&compress]{natbib}

\usepackage[stable]{footmisc}

\title{Testing gravity with $E_G$: mapping theory onto observations}

\author[a]{C. Danielle Leonard}
\emailAdd{danielle.leonard@physics.ox.ac.uk}

\author[a]{Pedro G. Ferreira}
\emailAdd{p.ferreira1@physics.ox.ac.uk}

\author[b]{Catherine Heymans}
\emailAdd{heymans@roe.ac.uk}

\affiliation[a]{Astrophysics, University of Oxford, Denys Wilkinson Building, Keble Road, Oxford, OX1 3RH, UK}
\affiliation[b]{Institute for Astronomy, University of Edinburgh, Royal Observatory, Blackford Hill, Edinburgh, EH9 3HJ, UK}

\abstract{We present a complete derivation of the observationally motivated definition of the modified gravity statistic $E_G$. Using this expression, we investigate how variations to theory and survey parameters may introduce uncertainty in the general relativistic prediction of $E_G$. We forecast errors on $E_G$ for measurements using two combinations of upcoming surveys, and find that theoretical uncertainties may dominate for a futuristic measurement. Finally, we compute  predictions of $E_G$ under modifications to general relativity in the quasistatic regime, and comment on the pros and cons of using $E_G$ to test gravity with future surveys.}

\notoc

\begin{document}

\maketitle

\section{Introduction}

Since the discovery of the accelerated expansion of the universe \cite{Riess1998, Perlmutter1999}, understanding the physical cause of this phenomenon has been an active area of cosmological interest. Several explanations have been proposed, including the standard cosmological constant, and the introduction of a dynamical dark energy component \cite{Zlatev1999, Carroll1998, Armendariz2001, Babichev2008}. Taking a different approach, it is possible to attribute accelerated expansion to modifications to the nature of gravity on cosmological scales \cite{gravrev}.  

In order to detect or constrain deviations from general relativity (GR), one normally focuses on the clustering properties of matter, with weak lensing, redshift-space distortions, and galaxy clustering widely recognised as powerful probes \cite{Schmidt2008, Thomas2009, Amendola2008, Tsujikawa2008, Zhang2007, Bertschinger2008, Knox2006, Ishak2006, Beutler2014, Raccanelli2013, Jennings2012}. The use of any single observation is typically subject to degeneracies and biases and so, for optimal constraints, observables are considered in combination. This may be via standard likelihood methods, as in for example \cite{Simpson2012}. Alternatively, observables can be strategically combined into a single statistic. This is the goal behind the statistic $E_G$.

$E_G$ was first proposed in 2007 \cite{Zhang2007} as a method of detecting deviations from GR while circumventing potential degeneracies with the galaxy bias. The suggested estimator involved a ratio between $C_{\kappa g}(l)$, the on-sky cross-spectrum of galaxy positions and  weak lensing convergence, and $P_{v g}(k)$, the cross-spectrum of velocities with galaxy positions, which could, in principle, be obtained from redshift-space distortion measurements. The same population of galaxies would be used to make both measurements, allowing the galaxy bias to cancel. Additionally, the required lensing measurement would be from galaxy-galaxy lensing, which is less susceptible to weak lensing systematic effects than cosmic shear.

The first measurement of $E_G$ was made three years later \cite{Reyes2010}, with a second measurement following some five years after that \cite{Blake2015}. Both have shown consistency with GR, as well as with other theories of gravity. These two measurements have employed a slightly different definition of $E_G$, first set out in \cite{Reyes2010}. This observationally-motivated definition replaces Fourier-space quantities with real-space equivalents, and substitutes the difficult-to-measure $P_{v g}(k)$ with the product of a projected galaxy correlation function and $\beta = f / b$ (where $f$ is the linear growth rate of structure and $b$ is the bias of the galaxy tracer). At linear scales, and under the assumption of constant bias, it has been assumed that the new observational definition should agree with the definition of $E_G$ as proposed in \cite{Zhang2007}.

In order to use $E_G$ to constrain or detect deviations from GR, we require predictions of $E_G$, for GR and for alternative theories of gravity. Some such predictions for the original definition of $E_G$ were made in \cite{Zhang2007}, and became the standard choice for comparison with measurement. However, the observational definition of $E_G$ is expected to agree with the original definition only under the above-mentioned assumption of constant bias. Additionally, as we will discuss, the observational definition contains several additional non-gravitational parameters, which may be degenerate with a departure from GR. It is the difference between the original proposal of \cite{Zhang2007} and the statistic that is actually used in data analysis that we wish to explore in this paper.

The remainder of this work is structured as follows. In Section \ref{section:deriv}, we review the original and the observational definitions of $E_G$, and derive an expression for the observational definition in terms of theoretical quantities in general relativity. In Section \ref{section:variables}, we consider how predictions for $E_G$ are affected by the variation of new parameters within the observationally-motivated definition, and we compare the associated uncertainties with the expected level of error on $E_G$ from two possible next-generation measurements. In Section \ref{section:modg}, we extend the derivation of Section \ref{section:deriv} to provide a full theoretical expression for $E_G$ under the $(\mu, \, \Sigma)$ parameterisation of modifications to GR in the quasistatic regime. We use this expression to compute $E_G$ under deviations from GR, and discuss how well future surveys may constrain alternative theories of gravity using $E_G$. We conclude in Section \ref{section:conc}.

\section{Derivation of a theoretical expression for $E_G(R)$}
\label{section:deriv}

\subsection{Definitions and review}
\label{subsection:defn}
\noindent
We will use the scalar-perturbed Friedmann-Robertson-Walker metric with the following convention:
\begin{equation}
ds^2=-(1+2\Psi)dt^2+a^2(1-2\Phi)d\vec{x}^2,
\label{scalarfrw}
\end{equation}
such that in GR, and in the absence of anisotropic stress, $\Phi=\Psi$.

For the remainder of the work, we refer to the original definition of $E_G$ set out in \cite{Zhang2007} as $E_{G}^0(l)$, and to the observationally-motivated definition underlying the statistic used in \cite{Reyes2010} as $E_G(R)$. We now reproduce the definitions of these quantities here. Note that both $E_G^0(l)$ and $E_G(R)$ are redshift-dependent, but we typically suppress this dependence for clarity.

First, $E_{G}^0(l)$ is defined as the expectation value of
\begin{equation}
\hat{E}_{G}^{0}(l, \Delta l)=\frac{C_{\kappa g}(l, \Delta l)}{3H_{0}^2a^{-1}\sum\limits_{\alpha} j_{\alpha}(l, \Delta l)P^{\alpha}_{vg}},
\label{E_orig}
\end{equation}
where $l$ is the magnitude of the two-dimensional on-sky Fourier-space wavenumber, $C_{\kappa g}(l, \Delta l)$ is the on-sky cross-spectrum of convergence and galaxy positions in bins of $\Delta l$, $P^{\alpha}_{vg}$ is the cross-spectrum of velocities and galaxy positions between $k_{\alpha}$ and $k_{\alpha+1}$, and $j_{\alpha}(l, \Delta l)$ is a weighting function which converts $P^{\alpha}_{vg}$ to an angular power spectrum. The expectation value of $\hat{E}_{G}^0(l, \Delta l)$ is then given by:
\begin{equation}
E_{G}^0(l)=\left[\frac{\nabla^2(\Psi+\Phi)}{3H_0^2a^{-1}f\delta_M}\right]_{k=l/\bar{\chi},\bar{z}}
\label{egexp}
\end{equation}
where $f$ is the linear growth rate of structure, $\delta_M$ is the matter overdensity field, and $\bar{\chi}$ is the comoving distance corresponding to redshift $\bar{z}$. The theoretical value of $E_G^0(l)$ would in fact be expected to be independent of $l$ for any theory of gravity which can be represented by a scale-independent generalised Newtonian constant, with a scale-independent relationship between $\Phi$ and $\Psi$. This is of course the case for GR, where it can be shown that $E_{G}^{0}(l)=\frac{\Omega_M(z=0)}{f(z)}$ \cite{Zhang2007}.

Next, the observationally-motivated $E_G(R)$ is defined in \cite{Reyes2010} to be:
\begin{equation}
E_G(R)=\frac{\Upsilon_{gm}(R)}{\beta \Upsilon_{gg}(R)}
\label{Eg(R)}
\end{equation}
where $\beta=f(z)/b$ ($b$ the galaxy bias as measured on linear scales) and $R$ is the transverse distance from the lens galaxy. $\Upsilon_{gm}(R)$ and $\Upsilon_{gg}(R)$ are annular differential surface densities (ADSDs) \cite{Baldauf2010}: modified versions of standard correlation functions which exclude information on scales below a cut-off threshold denoted $R_0$. 

Equation \ref{Eg(R)} is qualitatively similar to equation \ref{E_orig}. Both $\Upsilon_{gm}(R)$ and $C_{\kappa g}(l, \Delta l)$ represent a cross-correlation between galaxies and matter (with convergence, $\kappa$, mapping matter). Similarly, because $f\delta_M = v$ on linear scales, $\beta \times \Upsilon_{gg}(R)$ can be interpreted on those scales as a cross-correlation between galaxies and galaxy velocities, much like $P_{v g}(k)$. 

Specifically, $\Upsilon_{gm}(R)$ and $\Upsilon_{gg}(R)$ are defined as:
\begin{align}
\Upsilon_{gm}(R)&=\frac{2}{R^2}\int_{R_0}^{R}dR' R' \Sigma_{gm}(R')
    - \Sigma_{gm}(R)+\left(\frac{R_{0}}{R}\right)^2\Sigma_{gm}(R_0)
  \nonumber \\
&=\Delta \Sigma_{gm}(R)-\left(\frac{R_0}{R}\right)^2\Delta \Sigma_{gm}(R_0)
\label{ygm}
\end{align}
\begin{align}
\Upsilon_{gg}(R)&=\rho_c \Bigg[\frac{2}{R^2}\int_{R_0}^{R} dR' R'
  w_{gg}(R')-w_{gg}(R)+\left(\frac{R_0}{R}\right)^2 w_{gg}(R_0)\Bigg]
\label{ygg}
\end{align}
where $\rho_c$ is the critical density of the universe, $\Sigma_{gm}(R)$ is the projected surface mass density at $R$ (see, for example, equation 5 of \cite{Baldauf2010}), and $w_{gg}(R)$ is the projected galaxy autocorrelation function, as defined below in equation \ref{wgg}. $\Delta \Sigma_{gm}(R)$ is the excess differential surface mass density, given by:
\begin{equation}
\Delta \Sigma_{gm}(R) = \overline{\Sigma}_{gm}(R) - \Sigma_{gm}(R),
\end{equation}
where $\overline{\Sigma}_{gm}(R)$ is the average value of $\Sigma_{gm}$ within a circle of radius $R$.

\subsection{Deriving $E_G(R)$ in general relativity}
\label{subsection:EgDeriv_gr}
\noindent
Having now reviewed the definitions of both $E_{G}^0(l)$ and $E_G(R)$, we endeavour to develop a full theoretical expression for $E_G(R)$. In this section, we do so in the context of GR; for an extension to alternative theories of gravity, see Section \ref{section:modg}. 

We begin by examining equation \ref{ygm} for $\Upsilon_{gm}(R)$. There, $\Upsilon_{gm}(R)$ is seen to be directly dependent upon $\Delta\Sigma_{gm}(R)$, which is in turn related to the tangential shear observed at a transverse distance $R$ from a lens, $\gamma_t^g(R)$. In the case where all lenses are assumed to lie in a thin distribution at $z=z_l$, this relationship is given by: 
\begin{align}
\Delta \Sigma_{gm}(z_l, R) &= \langle \gamma_t^g(R) \rangle \left[ \int_{z_l}^\infty dz' \frac{P_s(z')}{\Sigma_{c}(z_l,z')} \right]^{-1},
\label{DeltaSig_narrowlens_anysource}
\end{align}
where $P_s(z)$ is the normalised redshift distribution of sources and $\Sigma_{c}$ is the critical surface mass density, given by \cite{Blake2015}:
\begin{equation}
\Sigma_{c}(z,z')=\frac{c^2\chi(z')}{4\pi G (\chi(z') - \chi(z))\chi(z)(1+z)}.
\label{SigcritRCS}
\end{equation}
It will be convenient to introduce the quantity $\overline{\Sigma_c^{-1}}(z_l)$, which is defined as:
\begin{equation}
\overline{\Sigma_c^{-1}}(z_l) = \int_{z_l}^\infty dz' \frac{P_s(z')}{\Sigma_{c}(z_l,z')}.
\label{barSig}
\end{equation}
In this work, since we will employ only hypothetical source redshift distributions which are defined apriori, we assume that source redshifts are perfectly known. The result is that for a single lens/source pair, we have:
\begin{equation}
\overline{\Sigma_c^{-1}}(z_l, z_s)= \Bigg\{ \begin{array}{cc} \Sigma_c^{-1}(z_l, z_s) &, z_s > z_l \\
0 &, z_s \le z_l 
\end{array},
\label{barSig2}
\end{equation}
which results in the following expression for $\Delta \Sigma_{gm}(R, z_l, z_s)$:
\begin{equation}
\Delta \Sigma_{gm}(R, z_l, z_s)= \Bigg\{ \begin{array}{cc} \gamma_t^g(R, z_l, z_s) \Sigma_c(z_l, z_s) &, z_s > z_l  \\
0 &, z_s \le z_l 
\end{array}.
\label{DeltaSigSimp}
\end{equation}
$\langle \gamma_t^g \rangle$ has become $\gamma_t^g$ in the limit of a single lens/source pair.

Extending to the full lens and source galaxy populations, the inclusion of redshift information becomes more complex. We have already introduced $P_s$, which characterises the source galaxy redshift distribution. For the lenses, we make the simplifying assumption of an effective redshift for the population, which we denote $\bar{z}_l$. There are a number of reasons for this choice, primary amongst them a reduction in computational difficulty. Throughout the work, we assume that all lenses are at $\bar{z}_l$, however, for a more general version of these expressions, see Appendix \ref{section:nozeff}. 

An estimator for $\Delta \Sigma_{gm}(R)$ is obtained by minimum-variance weighting each lens/source pair by a factor of $\overline{\Sigma_c^{-1}}^2$ \cite{Mandelbaum2005}, and then by summing over all lens/source pairs. We work with a continuous source galaxy distribution, and so replace summation with integration. Changing variables from redshift to comoving distance, the resulting expression is:
\begin{align}
\Delta\Sigma_{gm}(R)&=\frac{1}{\bar{w}} \int_0^{\chi_H} d\chi_s W_s(\chi_s)\overline{\Sigma_{c}^{-1}}(\bar{\chi}_l, \chi_s) \gamma_t^g(R, \bar{\chi}_l, \chi_s)
\label{deltasigthe1}
\end{align}
where $W_s(\chi)$ is the equivalent distribution to $P_s(z)$, and $\bar{w}$ is a normalisation accounting for the weighting factor:
\begin{align}
\bar{w}&= \int_0^{\chi_H} d \chi_s W_s(\chi_s)\left(\overline{\Sigma_{c}^{-1}}(\bar{\chi}_l, \chi_s) \right)^2.
\label{wbar2}
\end{align}

We now have an expression for $\Delta \Sigma_{gm}(R)$ in terms of the tangential shear. We know that $\gamma_t^g$ is related to the convergence $\kappa$ via (see, for example, \cite{MiraldaEscude1995}):
\begin{equation}
\gamma_t^g(R, \chi_s, \bar{\chi}_l)=-\frac{1}{2}\frac{d\bar{\kappa}(R, \chi_s, \bar{\chi}_l)}{d\ln(R)}
\label{gammatlogk}
\end{equation} 
and that the convergence in a direction ${\hat \theta}$ is given by (see, for example, \cite{Guzik2000}):
\begin{equation}
{\bar \kappa}({\hat \theta}, \chi_s)=\frac{3}{2}\left(\frac{H_0}{c}\right)^2\Omega_M(z=0)\int_0^{\chi_s}d\chi\frac{\chi(\chi_s-\chi)}{\chi_sa(\chi)}\delta_M(r(\chi{\hat \theta},\chi)),
\label{kappa_guzik}
\end{equation}
where $\chi_s$ is the comoving distance to the source in question. However, equation \ref{gammatlogk} requires not ${\bar \kappa}({\hat \theta}, \bar{\chi}_l, \chi_s)$, but $\bar{\kappa} (R, \bar{\chi}_l, \chi_s)$. To get this, we first modify equation \ref{kappa_guzik} to obtain the convergence in direction ${\hat \theta_2}$ given that there is a galaxy in direction ${\hat \theta_1}$ at comoving distance $\bar{\chi}_l$. We then invoke the small angle approximation, and change variables such that we replace the two directions $\hat{\theta}_1$ and $\hat{\theta}_2$ with the transverse on-sky distance between them, to find:
\begin{align}
\gamma_t^g(R, \chi_s, \bar{\chi}_l)&=-\frac{3}{2}\left(\frac{H_0}{c}\right)^2\Omega_M(z=0) \int_0^{\chi_s}d\chi \frac{\chi(\chi_s-\chi)}{a(\chi)\chi_s} \nonumber \\ &\times \frac{d}{d\ln R}\left[\int_{S_R} \frac{R'd R'}{R^2} \xi_{gm}(\sqrt{R'^2+(\chi-\bar{\chi}_l)^2}, \bar{z}_l, b)\right],
\label{gammat1}
\end{align}
where $\xi_{gm}$ is the real-space galaxy-matter cross-correlation function. We insert equations \ref{SigcritRCS} and \ref{gammat1} into equation \ref{deltasigthe1}, to find:
\begin{align}
\Delta&\Sigma_{gm}(R)=\frac{-\rho_c \Omega_M(z=0)}{\bar{w}}\int_0^{\chi_H} d\chi_s W_s(\chi_s)\frac{(4\pi G)}{c^2}\overline{\Sigma_{c}^{-1}}(\bar{\chi}_l, \chi_s) \int_0^{\chi_s}d\chi \frac {\chi(\chi_s-\chi)}{a(\chi)\chi_s} \nonumber \\ & \times \frac{d}{d\ln R}\left[\int_{S_R} \frac{R'd R'}{R^2} \xi_{gm}(\sqrt{R'^2+(\chi-\bar{\chi}_l)^2}, \bar{z}_l, b)\right].
\label{deltasigthe2}
\end{align}
It will be convenient to introduce a modified version of the source window function, into which we will absorb the piecewise nature of $\overline{\Sigma_c^{-1}}$ as given in equation \ref{barSig2}:
\begin{displaymath}
\overline{W}_s(\chi_s)= \Bigg\{ \begin{array}{cc} W_s(\chi_s) &, \chi_s > \bar{\chi}_l \nonumber \\
0 &, \chi_s \le \bar{\chi}_l 
\end{array}.
\end{displaymath}
Incorporating this, changing the order of integrals, and rewriting $\chi=\bar{\chi}_l+\Delta$ gives:
\begin{align}
\Delta&\Sigma_{gm}(R)=\frac{-\rho_c \Omega_M(z=0)}{\bar{w}}  \int d\Delta \frac{(4\pi G)^2}{c^4} \frac{ \bar{\chi}_l (\bar{\chi}_l+\Delta)}{a(\bar{\chi}_l) a(\bar{\chi}_l+\Delta)} \int_{\bar{\chi}_l+\Delta}^{\chi_H} d\chi_s \overline{W}_s(\chi_s) \nonumber \\ &\times (\chi_s-\bar{\chi}_l)  \frac{(\chi_s-\bar{\chi}_l-\Delta) }{\chi_s^2} \frac{d}{d\ln R}\left[\int_{S_R} \frac{R'd R'}{R^2} \xi_{gm}(\sqrt{R'^2+(\Delta)^2}, \bar{z}_l, b)\right].
\label{deltasigwDelta}
\end{align}
Inserting then equation \ref{deltasigwDelta} into equation \ref{ygm}, we find:
\begin{align}
\label{ygm2}
&\Upsilon_{gm}(R)= \frac{\rho_c \Omega_M(z=0)}{\bar{w}} \int d\Delta \frac{(4\pi G)^2 \bar{\chi}_l (\bar{\chi}_l+\Delta)}{c^4a(\bar{\chi}_l) a(\bar{\chi}_l+\Delta)} \int_{\bar{\chi}_l+\Delta}^{\chi_H} d\chi_s \overline{W}_s(\chi_s) \frac{(\chi_s-\bar{\chi}_l-\Delta)(\chi_s-\bar{\chi}_l)}{\chi_s^2} \nonumber \\
&\times  \Bigg[\frac{2}{R^2}\int_{R_0}^{R}R'dR' \xi_{gm}\left(\sqrt{R'^2+\Delta^2}, \bar{z}_l, b\right)-\xi_{gm}\left(\sqrt{R^2+\Delta^2}, \bar{z}_l, b\right) \nonumber \\ &+\left(\frac{R_0}{R}\right)^2\xi_{gm}\left(\sqrt{R_0^2+\Delta^2}, \bar{z}_l, b\right)\Bigg],
\end{align}
where we have used the fact that $R\frac{d}{dR}\left(\frac{2}{R^2}\int_0^R R'dR'F(R')\right)=2\left(F(R)-\frac{2}{R^2}\int_0^R R'dR'F(R')\right)$. 

We now move on to $\Upsilon_{gg}(R)$, the equivalent ADSD for galaxy clustering. Once again, we assume that the galaxy clustering measurement will have an effective redshift; this is standard practice in this case. $\Upsilon_{gg}(R)$ depends on $w_{gg}(R)$, the projected galaxy correlation function, given by (see, for example, \cite{Zehavi2005}):
\begin{equation}
w_{gg}(R)=\int_{-P}^{P} d\Delta \xi_{gg}\left(\sqrt{\Delta^2+R^2}, \bar{z}_l, b^2\right).
\label{wgg}
\end{equation}
$P$ here is a constant which determines the line-of-sight distance over which $\xi_{gg}$ is projected. We then insert equation \ref{wgg} into equation \ref{ygg}:
\begin{align}
\Upsilon_{gg}(R)&=\rho_C \int_{-P}^{P} d\Delta \Bigg[\frac{2}{R^2}\int_{R_0}^{R}R'dR' \xi_{gg}\left(\sqrt{R'^2+\Delta^2},\bar{z}_l, b^2\right)\nonumber \\ &-\xi_{gg}\left(\sqrt{R^2+\Delta^2},\bar{z}_l, b^2\right) +\left(\frac{R_0}{R}\right)^2\xi_{gg}\left(\sqrt{R_0^2+\Delta^2}, \bar{z}_l, b^2\right)\Bigg].
\label{ygg2}
\end{align}

The final element which is required to construct $E_G(R)$ is $\beta(z)=f(z) / b(z)$, given by solving the following differential equation (where $f(z) = \delta_M'(z) / \delta_M(z)$ and a prime represents a derivative with respect to $\log(a)$):
\begin{equation}
\delta_M'' + \left(1 + \frac{(aH)'}{aH}\right)\delta_M' - \frac{3}{2} \Omega_M \delta_M = 0
\label{fz}
\end{equation}
It is important to note that in $E_G(R)$, $b$ as it appears in $\beta$ is always scale-independent by definition, despite the fact that in general the bias may take a scale-dependent form. Because the goal of this work is to examine the observational definition of $E_G(R)$ as it is currently applied in practice, we preserve this feature, and always take $\beta$ as a scale-independent quantity. 

We have now developed a full theoretical expression for each of the constituents of $E_G(R)$, summarised in equations \ref{ygm2}, \ref{ygg2}, and \ref{fz}. We remind the reader that the above expressions are for the case where the lens galaxies are assumed to be at an effective redshift $\bar{z}_l$; the more general case is given in Appendix \ref{section:nozeff}. We see that these expressions contain several variable quantities which could affect predictions of $E_G(R)$: the distribution of source galaxies $W_s$, the projection length $P$, the cut-off scale $R_0$, and the bias $b$ (particularly if scale-dependent). We now examine the effect of varying these quantities on the predictions of $E_G(R)$ in GR.

\section{Understanding theoretical uncertainties}
\label{section:variables}
\noindent
Given now the expressions derived in the previous section, we have the means to compute predictions of $E_G(R)$ in GR. We take advantage of this to examine the effect of varying $W_s$, $P$, $b$ and $R_0$ on the general relativistic prediction. 

It is possible that in future measurements of $E_G(R)$, the uncertainty associated with variations to these parameters may be significant. In order to investigate this, we compare with forecast errors from two possible future measurements of $E_G(R)$: one next-generation measurement, and one more futuristic. These measurements are characterised as follows:
\begin{enumerate}
\item{{\bf DESI + DETF4:} Galaxies for $\Upsilon_{gg}(R)$ and $\beta$, as well as the lens galaxies for $\Upsilon_{gm}(R)$, are provided by the Dark Energy Spectroscopic Instrument (DESI) Luminous Red Galaxy sample \cite{Levi2013}. Source galaxies for the measurement of $\Upsilon_{gm}(R)$ come from a Dark Energy Task Force 4 (DETF4) type survey \cite{EuclidRedBook}. For this measurement we take $\bar{z}_l=0.8$, and we assume a sky area of $9000$ square degrees from DESI.}
\item{{\bf SKA2 + LSST:} Galaxies for $\Upsilon_{gg}(R)$ and $\beta$, as well as the lens galaxies in $\Upsilon_{gm}(R)$, are provided by a Stage 2 survey of HI galaxies by the Square Kilometre Array (SKA) \cite{Santos2015}. Source galaxies for the measurement of $\Upsilon_{gm}(R)$, come from a survey by the Large Synoptic Survey Telescope (LSST) \cite{LSSTScience}. We use the LSST source redshift distribution given in \cite{Hojjati2012}. For this measurement we take $\bar{z}_l=1.0$, and we assume a sky area of $30000$ square degrees from the SKA survey.}
\end{enumerate}
Allowing the lens galaxy population to take an effective redshift has implications for the forecast errors, as we must neglect all lensing information from any source galaxies situated below this effective redshift. The result is that the error bars displayed may be inflated, and should be viewed as an upper bound. For a full explanation of how observational errors are calculated, see Appendix \ref{section:variance}. Note that all error bars displayed are $1\sigma$.

We now consider the possible  uncertainty which may be introduced by varying $W_s$, $P$, $b$, and $R_0$, and compare with forecast errors. In doing so, we select a fiducial set of quantities to hold fixed while varying each of these parameters one at a time.  Following \cite{Reyes2010, Blake2015}, we let $R_0 = 1.5$ Mpc/h. The galaxy bias $b$ is set as constant. We set $P=500$ Mpc/h, motivated by the fact that, as we will see, it is a minimum value for which $E_G(R)$ is insensitive to the specific choice. Finally, we let $W_s$ be given by equation \ref{nofz} (below) for the DESI+DETF4 measurement, and by the similar form given in \cite{Hojjati2012} for the SKA2+LSST measurement.

The correlation function of matter, $\xi_{mm}$, is calculated using:
\begin{align}
\xi_{mm}(r)&=\frac{1}{2 \pi^2}\int_{0}^{\infty}dk k^2 P_{mm}(k, \bar{z}_{l})\frac{\sin(kr)}{kr}e^{-k^2}
\label{xi}
\end{align}
where $P_{mm}$ is the non-linear theoretical power spectrum, which we calculate using the publicly available code CAMB \cite{Lewis2000} with Halofit \cite{Smith2003} and the 2015 Planck parameters \cite{Planck2015}. Recall also that we have $P_{gg}(k)= b^2 P_{mm}(k)$ and $P_{gm}(k) = b P_{mm}(k)$, and so for constant galaxy bias, equation \ref{xi} can be trivially adapted to compute $\xi_{gm}$ and $\xi_{gg}$.

\subsection{Source galaxy redshift distribution: $W_s$}
\label{subsection:dndz_shape}
\noindent
First, we consider the effect of allowing the shape of the source galaxy distribution, characterised by $W_s(\chi)$, to vary. We consider three cases, all of which are designed as modifications to the DESI+DETF4 measurement scenario:
\begin{itemize}
\item{{\bf Case 1:} All sources are assumed to be at a single comoving distance: $W_s(\chi_s)=\delta(\chi_s-\bar{\chi}_s)$. We set $\bar{\chi}_s$ such that all sources are at redshift $z=0.9$. This is an idealised scenario, which we do not suggest to represent reality, but rather to act as a limiting case in which we should recover the GR prediction put forth in \cite{Zhang2007}.
}
\item{{\bf Case 2:} The source distribution is modelled by a normalised Gaussian in $\chi$, such that the mean of the Gaussian is at $z=0.9$, and the square root of the variance given by $\sigma_s=60$ Mpc/h.
}
\item{{\bf Case 3:} The source distribution is given by \cite{Smail1994}:
\begin{equation}
W_s(\chi_s) = \frac{z(\chi_s)^\alpha}{N_0} e^{-\left(\frac{z(\chi_s)}{z_0}\right)^\beta} \frac{H(\chi_s)}{c},
\label{nofz}
\end{equation}
where $N_0$ is a normalisation factor such that $\int d\chi_s W_s(\chi_s)=1$. We set $\alpha=2$, $\beta=1.5$, and $z_0=0.9/1.412$, in agreement with a DETF4 type survey (see, for example, \cite{Thomas2009}). We normalise between $z=0.5$ and $z=2$, and set $W_s(\chi_s)$ to $0$ outside of this range.}
\end{itemize}

Calculating $E_G(R)$ for these three cases, we find that $E_G(R) = \frac{\Omega_M(z=0)}{f(z=0.9)}$ within $1\%$, for the range $R=2-80$ Mpc/h. This suggests that variation in the source galaxy distribution shape has no effect on predictions of $E_G(R)$. 

We pause to consider a key difference between $E_G(R)$ and $E_G^0(l)$, and the related implications of this result. $E_G^0(l)$, defined in Fourier-space, assumes the inclusion of only linear scales. However, $E_G(R)$ is defined in real space, where scales are not so easily separated. It is therefore possible that the mere inclusion of non-linearities in the correlation function would cause $E_G(R)$ to deviate from the expected general relativistic value at small scales. The current result demonstrates that this is not the case, at least for the fiducial parameter set that we have selected. This can be attributed to the fact that nonlinearities enter into $\Upsilon_{gm}(R)$ and $\Upsilon_{gg}(R)$ via the same combination of correlation function terms, as seen in equations \ref{ygm2} and \ref{ygg2}, and therefore effectively cancel out. 

\subsection{Projection length: $P$}
\noindent
We now consider the consequences of varying the projection length $P$. There are two related theoretical effects which could alter the value of $E_G(R)$ in this scenario:
\begin{itemize}
\item{The projection may be over an insufficiently large line-of-sight separation to allow the galaxy-galaxy correlation to become negligible over this distance, thus losing three-dimensional power in the projection.}
\item{Although we have expressed $w_{gg}(R)$ in terms of the real-space correlation function, in reality the three-dimensional correlation function is measured in redshift-space. If $P$ is insufficiently large, the effect of measuring in redshift-space will be non-negligible, as not all redshift-space effects will cancel along the line-of-sight.}
\end{itemize}
In order to examine these effects, we replace $\xi_{gg}(\sqrt{R^2+\Delta^2})$, in real-space, with $\xi^s_{gg}(R, \Delta)$, in redshift-space, given by:
\begin{align}
\xi^s_{gg}(R, \Delta) &= \frac{1}{4\pi^2}\int_{-\infty}^{\infty} d k_{||} \cos(k_{||} \Delta)  \int_0^{\infty} d k_{\perp} k_{\perp} J_0(k_{\perp} R) P^s_{mm}(k_{||}, k_{\perp}).
\label{xis}
\end{align}
We additionally incorporate Finger-of-God effects and a smoothing scale via:
\begin{equation}
P^s_{mm}(k_{||}, k_{\perp}) = \frac{(1+\beta \mu^2)^2 P_{mm}(k)e^{-k^2}}{1+(k\mu\sigma)^2}
\label{Psk}
\end{equation}
where $\sigma$, the Finger-of-God parameter is set to $400$ km/s, or $4/ \sqrt{2}$ in units of h/Mpc (see, for example, \cite{Cabre2009}). $k$ here is $\sqrt{k_{||}^2 + k_\perp^2}$.

\begin{figure*}[t]
\begin{center}
\includegraphics[width=\linewidth]{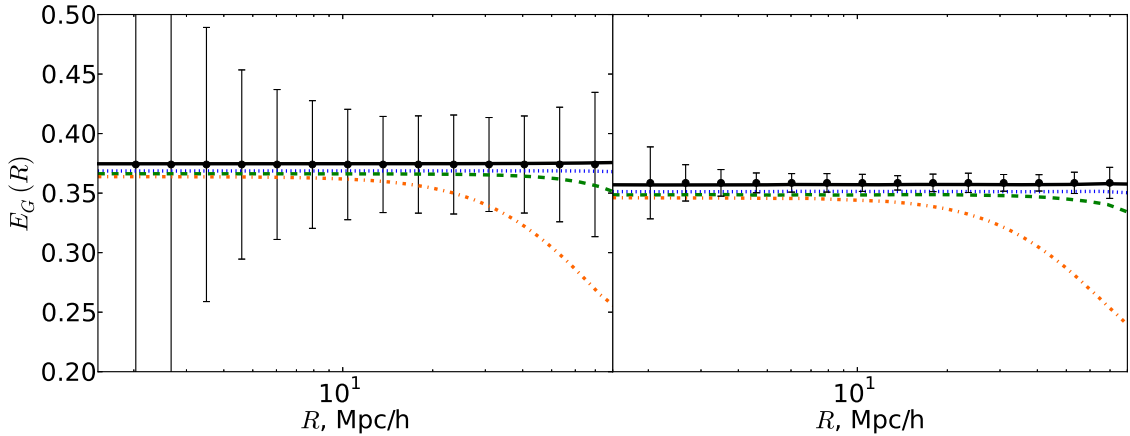}
\end{center}
\caption{Here we show how $E_G(R)$ is affected by varying the value of the projection length $P$. The left hand panel is for the case of a DESI+DETF4 measurement, with the right hand panel showing SKA2+LSST. We show predictions for values of $P=500$ (solid black), $P=250$ (dotted blue), $P=150$ (dashed green), and $P=50$ (dot-dashed orange). The solid black line also corresponds to the theoretical value of $\Omega_M(z=0) / f(\bar{z}_l)$. All values of $P$ are in Mpc/h.}
\label{figure:projection}
\end{figure*}

We compute $E_G(R)$ for a range of $P$ values, for both a DESI+DETF4 type measurement, and an SKA2+LSST type measurement. Figure \ref{figure:projection} shows the results. We see that for both measurement scenarios, when $P=500$ Mpc/h, $E_G(R)$ is given by the expected $\Omega_M(z=0) / f(\bar{z})$. However, as $P$ is reduced, $E_G(R)$ changes in two way: its overall value is reduced, and it is further suppressed at larger $R$. 

The suppression of $E_G(R)$ (or, equivalently, the boosting of $w_{gg}(R)$), can be understood as a result of the two factors mentioned above. First, reducing $P$ within this regime is equivalent to neglecting separations for which the correlation function is negative, and hence the value of the projected correlation function is incorrectly inflated. This effect is enhanced at larger $R$, where for fixed $P$, more of the negative region of the correlation function will be neglected than for smaller $R$. Second, redshift-space distortions result from peculiar velocities, which arise due to increased clustering and hence correspond to a boosted correlation function. Projecting over sufficiently long $P$ induces a cancellation of this effect along the line-of-sight, but as $P$ is reduced, this cancellation is negated and $w_{gg}(R)$ is boosted

As seen in Figure \ref{figure:projection}, for the DESI+DETF4 case, the effect of a reduction in the overall value of $E_G(R)$ at all scales is insignificant compared with the expected errors. The effect of a deviation from the expected value at large $R$ is more worrisome, but only for the smallest $P$ value which we plot. On the other hand, for the SKA2+LSST case, we see that both effects are significant in comparison with errors and would be important to account for in such a measurement.

Although it may seem from this result that it is optimal to choose the largest possible projection length, we note that this may not be the case. In practice, choosing a long projection length may increase the noise of the measurement, by incorporating redundantly large line of sight separations at which any structure is uncorrelated \cite{Zehavi2005}. This can be seen explicitly in the expression for the error of $\Upsilon_{gg}(R)$ derived in Appendix \ref{section:variance}. In reality, the choice of projection length must be balanced between theoretical uncertainty and shot noise.

\subsection{Scale-dependent bias}
\label{section:bias}
\noindent
The next possible source of theoretical uncertainty which we consider is scale-dependent galaxy bias. In this scenario, we can no longer trivially compute $\xi_{gg}$ and $\xi_{gm}$ from $\xi_{mm}$. We instead use: 
\begin{align}
\xi_{gm}(r)&=\frac{1}{2 \pi^2}\int_{0}^{\infty}dk k^2 b(k,\bar{z}_l) P_{mm}(k, \bar{z}_{l})\frac{\sin(kr)}{kr}e^{-k^2} \nonumber \\ 
\xi_{gg}(r)&=\frac{1}{2 \pi^2}\int_{0}^{\infty}dk k^2 b^2(k,\bar{z}_l) P_{mm}(k, \bar{z}_{l})\frac{\sin(kr)}{kr}e^{-k^2}
\label{xi_bias}
\end{align}
where $b(k,\bar{z}_l)$ is the galaxy bias, and $P_{mm}(k, \bar{z}_{l})$ is, for the bias models which we consider, the linear power spectrum.

We investigate the variation to $E_G(R)$ under two scale-dependent bias models. The first was introduced as part of the analysis for the 2dFGRS survey \cite{Cole2005}, and can be written as:
\begin{equation}
b_{\rm{2dF}}(k,z)=b_c(z)\sqrt{\left[\frac{1+Q(z)k^2}{1+A(z)k}\right]},
\label{colebias2}
\end{equation}
where $b_c(z)$, $A(z)$, and $Q(z)$ are constant at a given redshift and can be fit from n-body simulations. The second model features a power law in $k$, and can be described by \cite{Fry1993, Amendola2015}:
\begin{equation}
b_{\rm{PL}}(k,z)=b_0(z)+b_1(z)k^n,
\label{powerlawbias2}
\end{equation}
where once again $b_0(z)$ and $b_1(z)$ are $z$-dependent constants \cite{Amendola2015}, and $k$ is in units of $k_1=$ 1 h/Mpc. 

\begin{figure*}[t]
\begin{center}
\includegraphics[width=\linewidth]{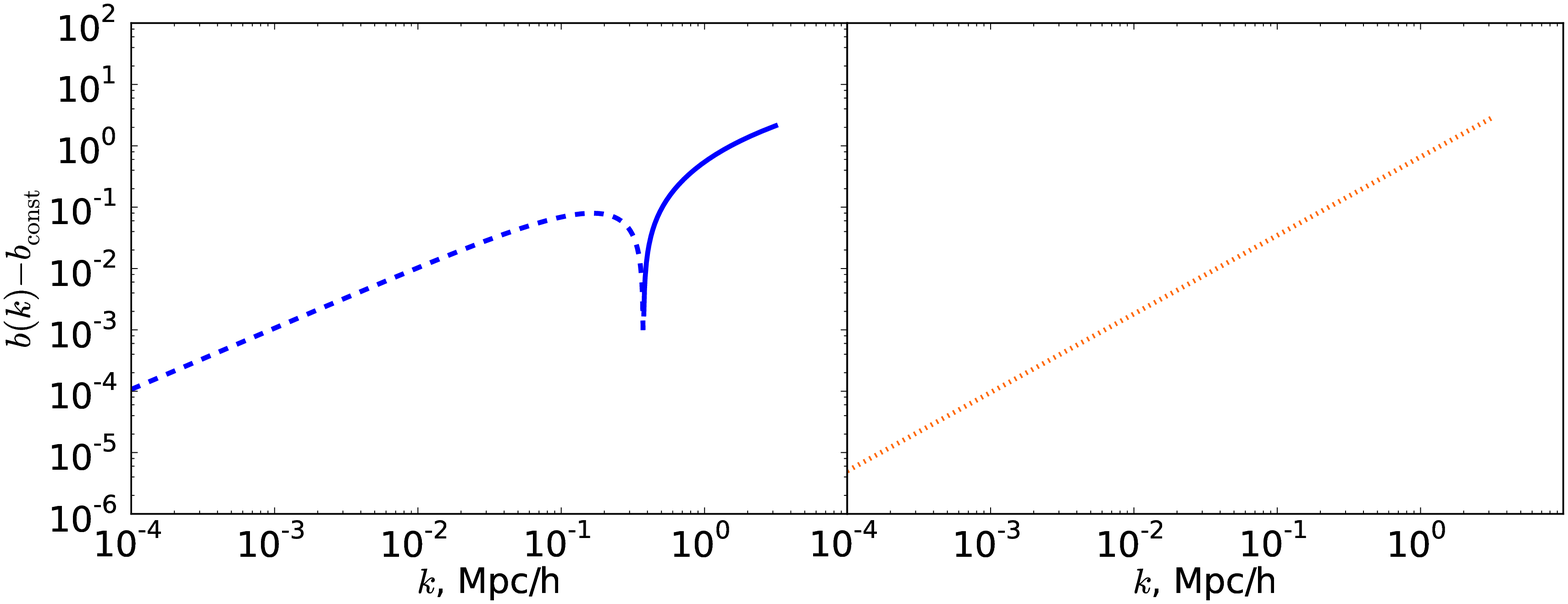}
\end{center}
\caption{The difference between $b(k)$ and the comparable constant bias value for the two scale-dependent bias models in the text (see equations \ref{colebias2} and \ref{powerlawbias2}), for the parameters given in \cite{Amendola2015} and at $z=0.8$ (as in a DESI+DETF4 measurement of $E_G(R)$). The left hand panel shows $b_{\rm{2dF}}(k)-b_c$, where the dashed portion of the curve is in reality negative, but is displayed in absolute value. The right hand panel shows $b_{\rm{PL}}(k)-b_0$. }
\label{figure:biasforms_show}
\end{figure*}

We compute $E_G(R)$ under these scale-dependent bias models, with other parameters fixed to fiducial values. The bias model parameters are taken to have values given in \cite{Amendola2015} for the case of $\bar{z}_l=0.8$ (for DESI+DETF4), and $\bar{z}_l=1.0$ (for SKA2+LSST), with $n=1.28$ for the power law bias model in both cases. For reference, we display both choices of bias as a function of scale in Figure \ref{figure:biasforms_show}, for the representative case of $\bar{z}_l=0.8$. As noted earlier, the factor of the bias which is introduced via $\beta(z)=f(z) / b(z)$ is independent of scale by definition, and should be chosen as $b_c(z)$ or $b_0(z)$ as appropriate.

\begin{figure*}[t]
\begin{center}
\includegraphics[width=\linewidth]{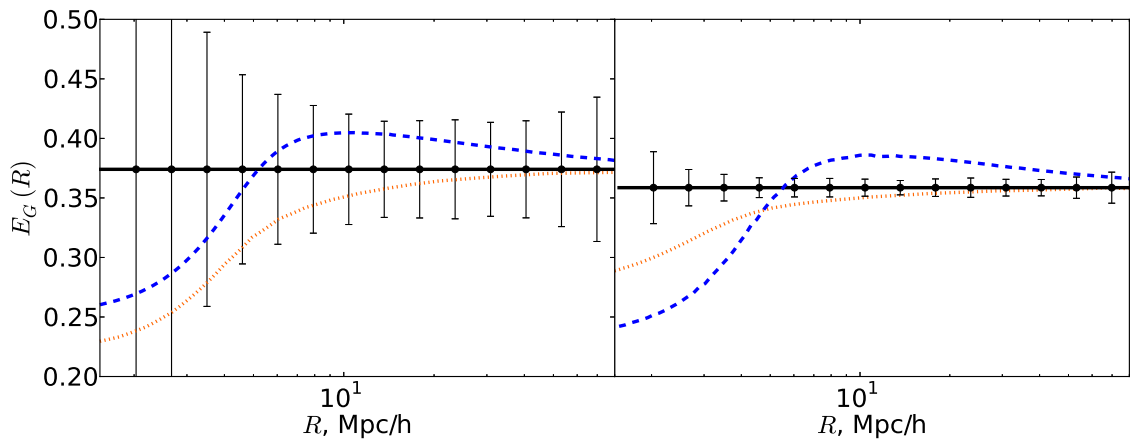}
\end{center}
\caption{$E_G(R)$ for the scale-dependent bias models described in the text. The left hand panel shows the case of a DESI+DETF4 measurement, and the right hand shows the case of SKA2+LSST.  Constant bias - black, solid. $b_{\rm{2dF}}$ (equation \ref{colebias2}) - blue, dashed. $b_{\rm{PL}}$  (equation \ref{powerlawbias2}) - orange, dotted.}
\label{figure:scaledepbias}
\end{figure*}

The resulting predictions are shown in Figure \ref{figure:scaledepbias}. We see that the inclusion of scale-dependent bias causes $E_G(R)$ to deviate considerably from $\frac{\Omega_M(z=0)}{f(\bar{z}_l)}$, particularly for small values of $R$. However, once again, for the DESI+DETF4 case, the effect is largely within the expected $1\sigma$ error, while for an SKA2+LSST measurement, this is not the case. The implication is that for a futuristic measurement of this type, scale-dependent bias must be accounted for. Generally, we would expect this to occur via accurate modelling of the scale-dependence. However, in the case of $E_G(R)$, we could imagine the possibility that an increase to the value of $R_0$ could sufficiently suppress the small-scale deviations from the expected value of $E_G(R)$. We explore this possibility in the next subsection.

Finally, we note that the predicted $E_G(R)$ resulting from the 2dFGRS bias model approaches $\frac{\Omega_M(z=0)}{f(\bar{z}_l)}$ less closely on this range than does the $E_G(R)$ prediction resulting from the power law bias model. One reason for this can be seen in Figure \ref{figure:biasforms_show}: $b_{\rm{2dF}}(k)$ deviates more significantly from constant at larger scales than does $b_{\rm{PL}}(k)$. This then translates to a more significant large-scale deviation of the value of $E_G(R)$ away from constant in the $b_{\rm{2dF}}(k)$ case.

\subsection{$R_0$}
\label{subsection:R0}
\noindent
The parameter $R_0$ defines the minimum value of $R$ from below which information is included in $E_G(R)$. In \cite{Reyes2010} and \cite{Blake2015}, $R_0$ was chosen to be $1.5$ Mpc/h, with \cite{Blake2015} showing their measurement to be unaffected by small variations to $R_0$. However, with the inclusion of scale-dependent bias, it is possible that the choice of $R_0$ may strongly influence the degree to which small scale information is removed. We therefore calculate $E_G(R)$ while varying $R_0$ under the scale-dependent bias models outlined in the previous section, as well as in the constant-bias case. Once again, we consider both a DESI+DETF4 and an SKA2+LSST type measurement.

\begin{figure*}[t]
\begin{center}
\includegraphics[width=\linewidth]{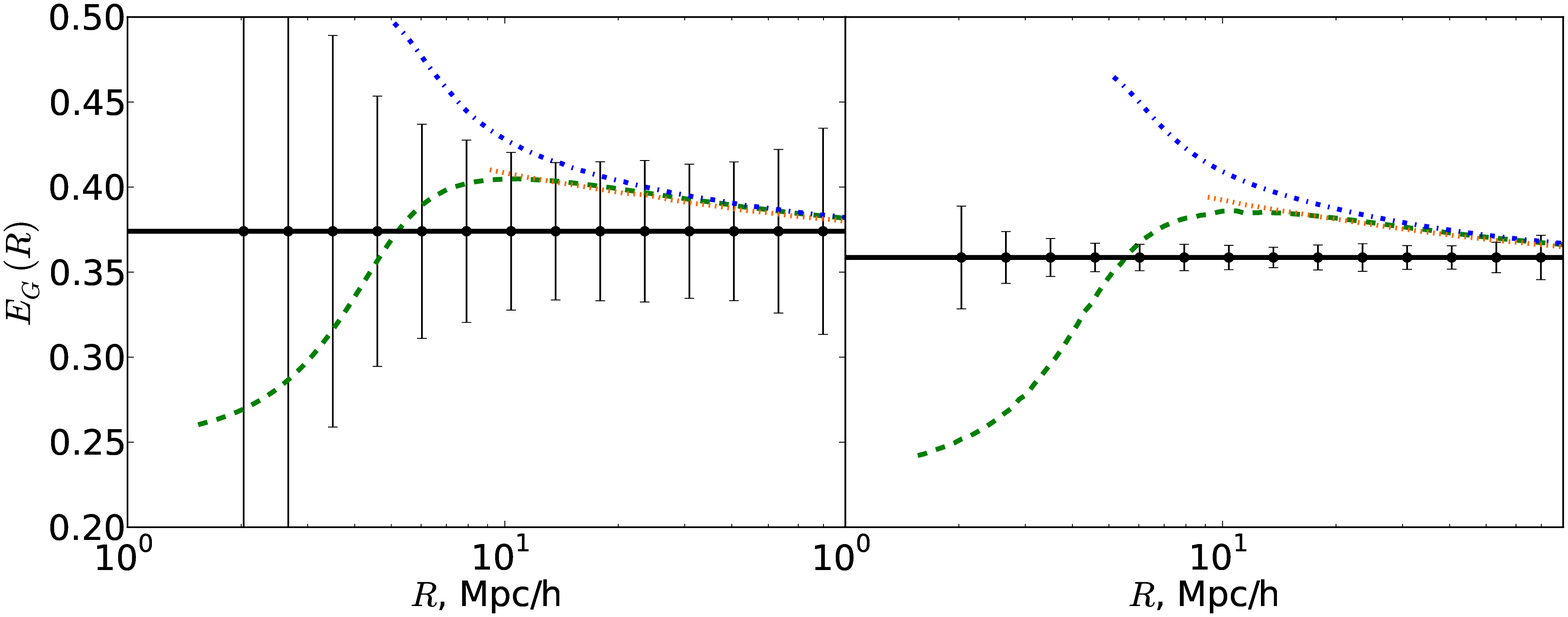}
\end{center}
\caption{$E_G(R)$ with variable $R_0$, for the case of $b_{\rm{2dF}}(k)$ (equation \ref{colebias2}). The left hand panel shows the case of a DESI+DETF4 measurement, and the right hand shows the case of SKA2+LSST. The curves displayed are for $R_0=1.5$ Mpc/h (green, dashed), $R_0=5.0$ Mpc/h (blue, dot-dashed), $R_0=9.0$ Mpc/h (orange, dotted). The solid black line shows the constant-bias case. Curves are shown in each case at values of $R$ greater than or equal to the relevant $R_0$.}
\label{figure:R0change_2dF}
\end{figure*}

\begin{figure*}[t]
\begin{center}
\includegraphics[width=\linewidth]{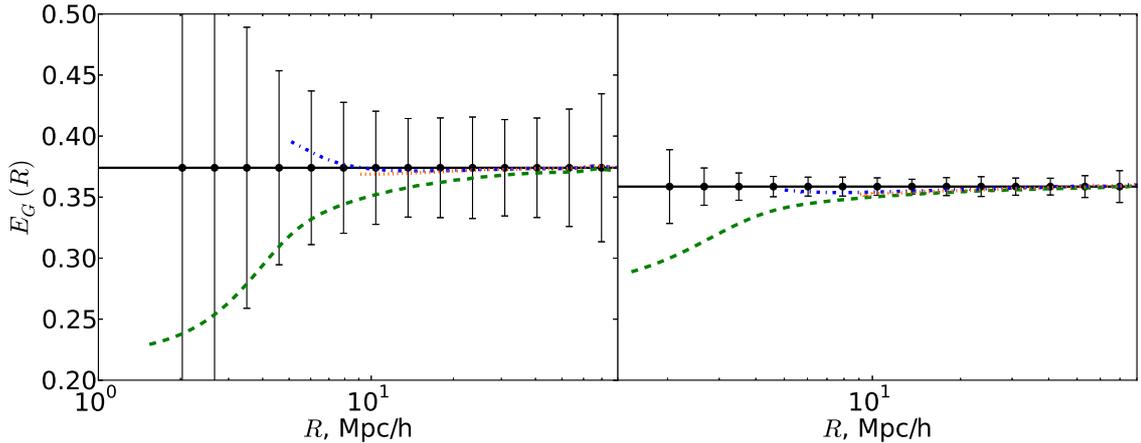}
\end{center}
\caption{$E_G(R)$ with variable $R_0$, for the case of the power-law bias model, given in equation \ref{powerlawbias2}. The left hand panel shows the case of a DESI+DETF4 measurement, and the right hand shows the case of SKA2+LSST. The curves displayed are for $R_0=1.5$ Mpc/h (green, dashed), $R_0=5.0$ Mpc/h (blue, dot-dashed), $R_0=9.0$ Mpc/h (orange, dotted). The solid black line shows the constant-bias case. Curves are shown in each case at values of $R$ greater than or equal to the relevant $R_0$.}
\label{figure:R0change_pl}
\end{figure*}

The result in the case of $b_{\rm{2dF}}(k)$ (given by equation \ref{colebias2}) is shown in Figure \ref{figure:R0change_2dF}, with Figure \ref{figure:R0change_pl} displaying the case of $b_{\rm{PL}}(k)$ (equation \ref{powerlawbias2}). Within each model, we plot $E_G(R)$ for $R_0=\{1.5, \, 5.0, \, 9.0\}$ Mpc/h. We see that for an SKA2+LSST measurement, increasing the value of $R_0$ will be insufficient to adequately excise information from small scales, particularly in the case of the 2dFGRS bias model. This implies  that increasing the value of $R_0$ will likely not prove a viable method of dealing with scale-dependent bias effects in the future, and that scale-dependent bias must be precisely modelled and understood if $E_G(R)$ is to be effectively used to test gravity. Even in the case of a DESI+DETF4 measurement, variation due to a change in $R_0$ is not fully contained within the $1\sigma$ error bars.

The effect of varying $R_0$ was also tested with the constant galaxy bias model. In this case, all choices of $R_0$ produced curves in which $E_G(R) \sim \frac{\Omega_M(z=0)}{f(\bar{z}_l)}$ on the range $R=2-80$ Mpc/h. For constant bias, the choice of $R_0$ therefore appears to have no effect on the predicted value of $E_G(R)$, as expected.

\section{Predictions of $E_G$ with modifications to general relativity}
\label{section:modg}
\noindent
We have now examined the degree to which the general relativistic prediction for $E_G(R)$ is affected by theoretical uncertainties, and how this may compare to the forecast errors for future measurements. However, we have not yet addressed the question of how these forecast errors compare to the expected change in $E_G(R)$ due to signatures of modified gravity. We now set out a theoretical expression for $E_G(R)$ under modifications to GR in the quasistatic regime. We then use this expression, along with the expressions for forecast errors derived in Appendix \ref{section:variance}, to comment on the level at which we might expect $E_G(R)$ to offer constraints on gravity from our two future measurement scenarios.

The quasistatic approximation states that within the range of scales that will be accessible to near-future cosmological surveys, time derivatives of new gravitational degrees of freedom can be set to zero (see, for example, \cite{Silvestri2013}). It has been shown \cite{Noller2014, Schmidt2009, Zhao2011, Barreira2013, Li2013} that within the regime of validity of this approximation, the most significant effects of a large class of cosmologically-viable alternatives theories of gravity can be captured by adding two functions of time and scale to the Poisson equation and the slip relation:
\begin{align}
2 \nabla^2 \Phi(\chi,k) &= 8 \pi G a(\chi)^2 \mu(\chi,k) \bar{\rho}_M\delta_M(\chi,k)\nonumber \\
\frac{\Phi(\chi,k)}{\Psi(\chi,k)}&=\gamma(\chi,k). 
\label{defgammu}
\end{align}
Note that dynamical quantities (such as $\delta_M$, $\Phi$, and $\Psi$) should be understood in the above equation as not necessarily evolving in the same way as they would in the GR case. That is, for example, $\delta_M \ne \delta_M^{GR}$ in general. 

We use this modified Poisson equation and slip relation to find expressions for $\Upsilon_{gm}(R)$ and $\Upsilon_{gg}(R)$ under modifications to GR. Motivated by previous work showing that any scale-dependence is expected to be sub-dominant \cite{BakerScales2014, Silvestri2013, Hojjati2014}, we restrict to the case of scale-independent $\mu$ and $\gamma$. The expression for $\bar{\kappa}$, given in equation \ref{kappa_guzik} for GR, then becomes:
\begin{align}
{\bar \kappa}({\hat \theta})&=\frac{3}{4}\left(\frac{H_0}{c}\right)^2\Omega_M(z=0)\int_0^{\chi_s}d\chi\left(1+\frac{1}{\gamma(\chi)}\right)\mu(\chi)\frac{\chi(\chi_s-\chi)}{\chi_sa(\chi)}\delta_M(r(\chi{\hat \theta},\chi)).
\label{kappa_guzik_MG}
\end{align}
Proceeding then in precisely the same manner as in the GR case, we obtain $\Upsilon_{gm}(R)$:
\begin{align}
\Upsilon_{gm}&(R)=\frac{\rho_c \Omega_M b}{2\bar{w}}\int d\Delta \frac{(4\pi G)^2 \bar{\chi}_l (\bar{\chi}_l+\Delta)}{c^4a(\bar{\chi}_l) a(\bar{\chi}_l+\Delta)}\int_{\bar{\chi}_l+\Delta}^{\chi_H} d\chi_s \overline{W}_s(\chi_s) w_s(\chi_s)  \nonumber \\ & \times   \left(1+\frac{1}{\gamma(\bar{\chi}_l+\Delta)}\right) \mu(\bar{\chi}_l+\Delta) (\chi_s-\bar{\chi}_l-\Delta)\frac{(\chi_s-\bar{\chi}_l)}{\chi_s^2} \nonumber \\& \times\Bigg[\frac{2}{R^2}\int_{R_0}^{R}R'dR' \xi^{MG}_{gm}\left(\sqrt{R'^2+\Delta^2}, \bar{z}_l, b\right)-\xi^{MG}_{gm}\left(\sqrt{R^2+\Delta^2}, \bar{z}_l, b\right) \nonumber \\ &+\left(\frac{R_0}{R}\right)^2\xi^{MG}_{gm}\left(\sqrt{R_0^2+\Delta^2}, \bar{z}_l, b\right)\Bigg].
\label{upgm_mg}
\end{align}
$\Upsilon_{gg}$ simply becomes:
\begin{align}
\Upsilon_{gg}&=\rho_C \int_{-P}^P d\Delta \Bigg[\frac{2}{R^2}\int_{R_0}^{R}R'dR' \xi^{MG}_{gg}\left(\sqrt{R'^2+\Delta^2},\bar{z}_l, b^2\right)\nonumber \\ &-\xi^{MG}_{gg}\left(\sqrt{R^2+\Delta^2},\bar{z}_l, b^2\right)+\left(\frac{R_0}{R}\right)^2\xi^{MG}_{gg}\left(\sqrt{R_0^2+\Delta^2},\bar{z}_l, b^2\right)\Bigg].
\label{ygg_mg}
\end{align}
$\xi_{gm}^{MG}$ and $\xi_{gg}^{MG}$ are computed using equation \ref{xi} where $P_{mm}(k, \bar{z}_l)$ is taken to be the matter power spectrum under the alternative theory in question. Finally, we also require the growth rate $f$ in terms of these quasistatic parameterising functions. It is given by solving an extended version of equation \ref{fz} \cite{Baker2014}:
\begin{equation}
\delta_M'' + \left(1 + \frac{(aH)'}{aH}\right)\delta_M' - \frac{3}{2}\Omega_M \delta_M \frac{\mu}{\gamma} = 0.
\label{fz_mod}
\end{equation}

In principle, we could now compute $E_G(R)$ under specific theories of gravity. However, it is common practice to pick a phenomenological form to model deviations from GR (see, for example, \cite{Simpson2012}). This method provides an attractive balance between ensuring the possibility of detecting deviations from GR and managing the analysis complexity. In order to allow for easy comparison with other work (for example, \cite{Thomas2009}), we follow this approach. We similarly choose to work in terms of the more common observational choice of parameterising functions $\{ \bar{\mu}, \, \Sigma\}$, instead of $\{ \mu, \, \gamma\}$. Under the assumption of small deviations from GR, and letting $\mu=1+\delta \mu$ and $\gamma=1+\delta\gamma$, these sets of functions are related by \cite{Leonard2015}:
\begin{align}
\Sigma&=\delta \mu- \frac{1}{2}\delta \gamma \nonumber \\
\bar{\mu}&= \delta \mu- \delta \gamma.
\label{paramfuncs}
\end{align}
We select a form of $\bar{\mu}(\chi)$ and $\Sigma(\chi)$ proposed in \cite{Ferreira2010} and employed in, for example, \cite{Simpson2012}:
\begin{align}
\bar{\mu}(\chi)&= \bar{\mu}_0\frac{\Omega^{GR}_{\Lambda}(\chi)}{\Omega^{GR}_{\Lambda}(\chi=0)} \nonumber \\ 
\Sigma(\chi)&= \Sigma_0\frac{\Omega^{GR}_{\Lambda}(\chi)}{\Omega^{GR}_{\Lambda}(\chi=0)}.
\label{latetimemugam}
\end{align}

Given equations \ref{upgm_mg}, \ref{ygg_mg} and \ref{fz_mod} in combination with equation \ref{paramfuncs} and \ref{latetimemugam}, we can now compute $E_G(R)$ for a range of $\bar{\mu}_0$ and $\Sigma_0$. In doing so, we will make several simplifying assumptions. First, because of the uncertain nature of nonlinear structure formation under general modifications to GR, we are unable to compute with confidence a nonlinear power spectrum, and therefore only linear-level effects are included. For this reason, it would be imprudent to attribute too much weight to our calculations below scales of roughly $10 $ Mpc/h. Additionally, we could in principle consider modifications to the expansion history as well as perturbative changes. However, due to the fact that the expansion rate has been observationally well-constrained to be near the $\Lambda$CDM quantity, and because many theories of interest are designed such that the $\Lambda$CDM expansion rate is reproduced, we consider here the case where the expansion history of the universe matches that of $\Lambda$CDM. We adopt the fiducial values of the previous section for the non-gravitational parameters, unless otherwise stated. 

\begin{figure*}[t]
\begin{center}
\includegraphics[width=\linewidth]{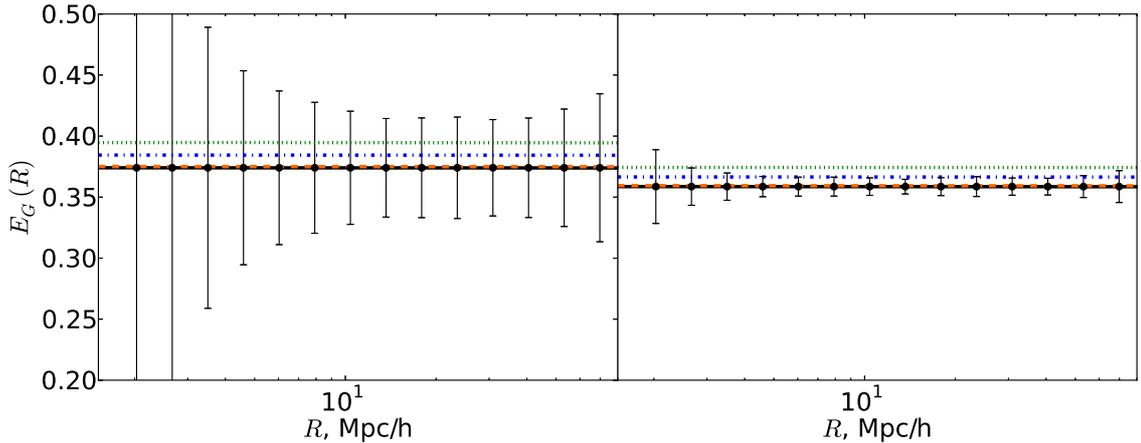}
\end{center}
\caption{$E_G(R)$ as a function of $R$ for different values of $\bar{\mu}_0$ and $\Sigma_0$. The left hand panel shows the case of a DESI+DETF4 measurement, and the right hand shows the case of SKA2+LSST.  $\bar{\mu}_0=\Sigma_0=0.01$ - orange, dashed; $\bar{\mu}_0=\Sigma_0=0.1$ - blue, dot-dashed; $\bar{\mu}_0=\Sigma_0=0.2$ - green, dotted; and for the GR case - black, solid.}
\label{figure:MG}
\end{figure*}

\begin{figure*}[t]
\begin{center}
\includegraphics[width=\linewidth]{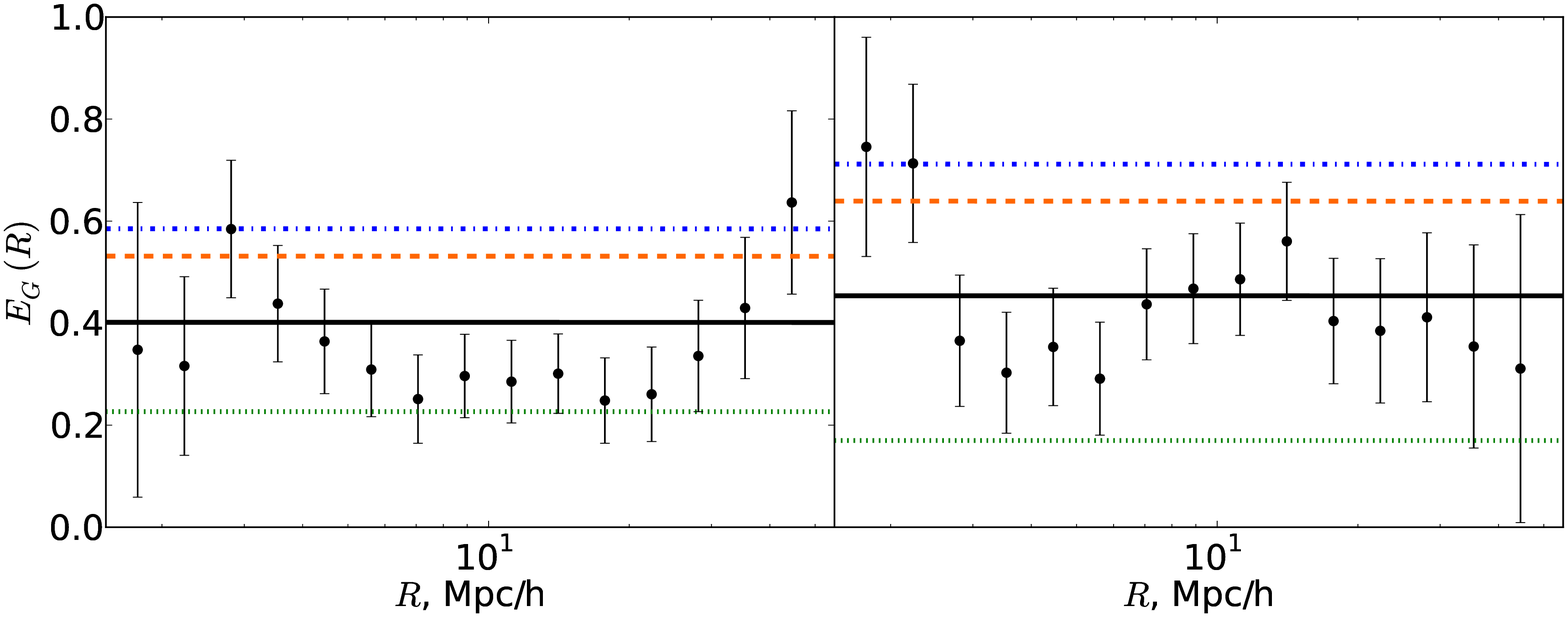}
\end{center}
\caption{$E_G(R)$ predictions as a function of $R$ for modifications to GR, shown with measurements of $E_G(R)$ from the RCSLenS collaboration \cite{Blake2015}. The left hand panel shows the measurement at $z=0.32$, and the right hand is at $z=0.57$. $\bar{\mu}_0=\Sigma_0=1.5$ - blue, dot-dashed; $\bar{\mu}_0=\Sigma_0=1.0$ - orange, dot-dashed, $\bar{\mu}_0=\Sigma_0=-1.0$ - green, dotted; and for the GR case - black, solid.}
\label{figure:MGBlake}
\end{figure*}

\begin{figure*}[t]
\begin{center}
\includegraphics[width=\linewidth]{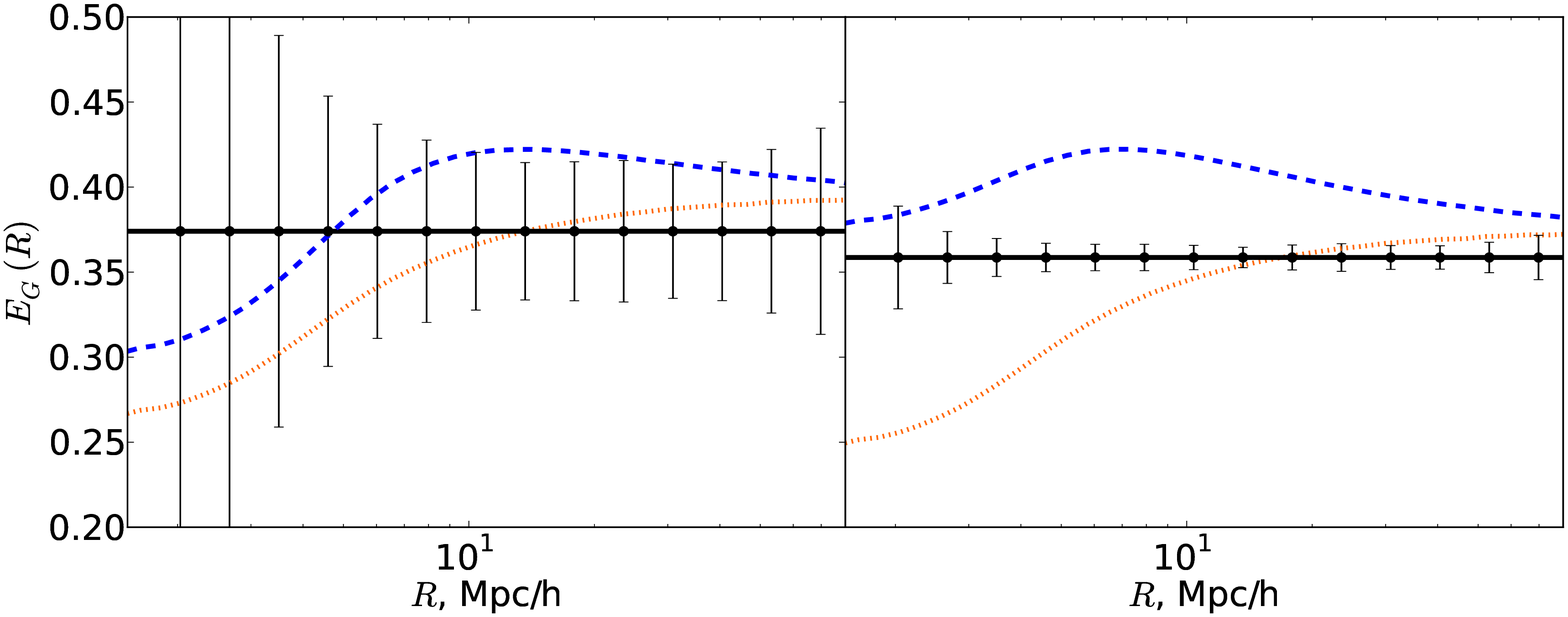}
\end{center}
\caption{$E_G(R)$ as a function of $R$ for $\bar{\mu}_0 = \Sigma_0=0.2$, with scale-dependent bias. The left hand panel shows the case of a DESI+DETF4 measurement, and the right hand shows the case of SKA2+LSST. Constant bias - black, solid. $b_{\rm{2dF}}$ (equation \ref{colebias2}) - blue, dashed. $b_{\rm{PL}}$  (equation \ref{powerlawbias2}) - orange, dotted.}
\label{figure:MG_otherparams}
\end{figure*}

The results of computing $E_G(R)$ under modifications to GR are shown in Figures \ref{figure:MG}, \ref{figure:MGBlake}, and \ref{figure:MG_otherparams}. In Figure \ref{figure:MG}, we see that, given the time-dependent form of $\bar{\mu}(\chi)$ and $\Sigma(\chi)$ that we have chosen, a measurement of $E_G(R)$ from DESI+DETF4 will struggle to distinguish a deviation from GR at a $20\%$ level in $\bar{\mu}_0$ and $\Sigma_0$. However, in the case of a measurement from SKA2+LSST, we expect to approach this level of $1\sigma$ constraint. In comparison, it has previously been shown that for a combination of cosmic shear, growth rate measurements, and baryon acoustic oscillations from a DETF4 type survey, constraints on $\bar{\mu}_0$ and $\Sigma_0$ are expected at a $5\%$ level \cite{Amendola2008, Leonard2015}. However, despite the looser constraint expected, $E_G(R)$ remains a valuable cross-checking mechanism, due to the different systematic effects acting on galaxy-galaxy lensing measurements. 

In order to compare these forecast errors with currently available measurements, we display in Figure \ref{figure:MGBlake} $E_G(R)$ at $z=0.32$ and $z=0.57$ as measured by the RCSLenS collaboration \cite{Blake2015}, in addition to the theoretical $E_G(R)$ at several choices of $\bar{\mu}_0$ and $\Sigma_0$. In this case, the chosen values of $\bar{\mu}_0$ and $\Sigma_0$ are not within the regime of small deviation from GR, so equation \ref{paramfuncs} is not used; rather the equivalent general expression is employed (see, for example, \cite{Simpson2012}). Fiducial parameters are as described above ($P=500$ Mpc/h, $R_0=1.5$ Mpc/h, and $b(k)$ is a constant), with the exception of $W_s$, which is given by equation \ref{nofz} with $\alpha=0.787$, $\beta=3.436$ and $z_0=1.157$ \cite{Benjamin2007}. We see that current $1\sigma$ constraints are at approximately the level of $|\bar{\mu}_0|=|\Sigma_0|=1.0$. These looser constraints are as expected, since these measurements are intended primarily as a `consistency check' for general relativity.

We next demonstrate the type of degeneracy which may exist between the gravitational parameters described here and the non-gravitational parameters of the previous section. To do so, we compute $E_G(R)$ for the case of $\bar{\mu}_0=\Sigma_0=0.2$, and this time allow the galaxy bias to take on scale-dependence as given in equations \ref{colebias2} and \ref{powerlawbias2}. The result is shown in Figure \ref{figure:MG_otherparams}, where it is clear that the scale-dependence of the bias may dominate over the effect of deviation from GR on all but the largest scales. This reinforces that notion that in order to achieve optimal constraints on gravity using next-generation measurements of $E_G(R)$, we will require accurate modelling of the galaxy bias.

\begin{figure*}[t]
\begin{center}
\includegraphics[width=\linewidth]{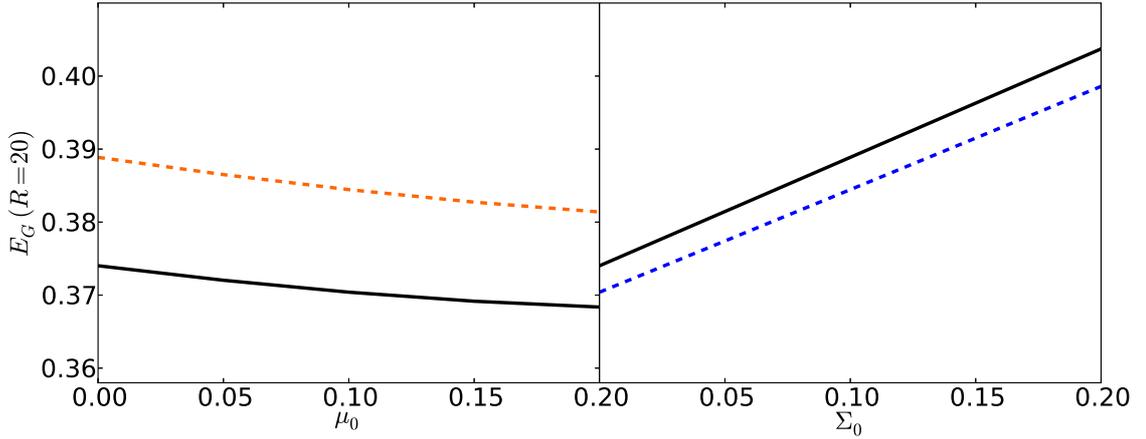}
\end{center}
\caption{$E_G$ at $R=20$ Mpc/h, as a function of $\bar{\mu}_0$ (left, with $\Sigma_0=0$ (black, solid) and $\Sigma_0=0.1$ (orange, dashed)) and as a function of $\Sigma_0$ (right, with $\bar{\mu}_0=0$ (black, solid) and $\bar{\mu}_0=0.1$ (blue, dashed)). The curves shown are for $\bar{z}_l=0.8$, as in the case of a DESI+DETF4 measurement of $E_G(R)$. The dependence of $E_G(R=20)$ on $\Sigma_0$ is clearly greater than that on $\bar{\mu}_0$.}
\label{figure:MG_mu_Sig_dep}
\end{figure*}

Finally, in order to better understand the dependence of $E_G(R)$ on $\bar{\mu}_0$ and $\Sigma_0$, we revert to a constant bias scenario, fix $R=20$ Mpc/h, and examine how $E_G(R=20)$ evolves as a function of $\bar{\mu}_0$ and of $\Sigma_0$. The results are seen in Figure \ref{figure:MG_mu_Sig_dep} for the representative case of $\bar{z}_l=0.8$. The left panel shows $E_G(R=20)$ as a function of $\bar{\mu}_0$ with $\Sigma_0=\{0, 0.1\}$, while in the right panel, $\Sigma_0$ varies and $\bar{\mu}_0=\{0, 0.1\}$. We see clearly that $E_G(R)$ exhibits a stronger dependence on $\Sigma_0$ than on $\bar{\mu}_0$. This can be understood by examining equation \ref{upgm_mg}. There, we see that $E_G(R)$ is primarily sensitive to the combination $\mu\left(1+\frac{1}{\gamma}\right)$, which, in the case of small deviations from GR, can be written as $2\left(\delta \mu - \frac{1}{2}\delta \gamma\right) = 2\Sigma $, as seen in equation \ref{paramfuncs}. $E_G(R)$ is affected by ${\bar{\mu}_0}$ primarily via the growth rate $f$ (because $\frac{\mu}{\gamma} \sim \delta \mu - \delta \gamma = \bar{\mu}$ in the case of small deviations from GR). However, changes to $f$ as a result of $\bar{\mu}_0$ are small (below $10\%$) at $z=0.8$. The relative insensitivity of $E_G(R)$ to $\bar{\mu}_0$ is an important limitation to bear in mind when selecting a method by which to detect or constrain deviations to GR on large scales. (Note the results shown in Figure \ref{figure:MG_mu_Sig_dep} are independent of the choice of $R$.)

\section{Conclusions}
\label{section:conc}
\noindent
Directly combining observables in order to circumvent potential degeneracies is an attractive prospect for testing gravity on cosmological scales. Here, we have examined in detail the observational definition of one such combination: $E_G(R)$. We have done so with the goal of understanding how parameters which were not present in the original version of the statistic may introduce theoretical uncertainty, and how this uncertainty may compare to forecast errors from two futuristic measurements. 

Our investigation reveals that of four potential sources of theoretical uncertainty (the distribution of source galaxies $W_s$, the projection length $P$, the cut-off scale $R_0$, and the bias $b$), only $W_s$ has no effect on the theoretical value of $E_G(R)$. The general relativistic prediction is sensitive to $P$, to the scale-dependence of the bias $b(k)$, and, under scale-dependent bias, to the value $R_0$ within which small scale information is discarded.  Until now, these few-percent changes of $E_G(R)$ away from $E_G^0(l)$ have not been critical when compared to statistical error \cite{Reyes2010, Blake2015}. Similarly, we find that in comparing theoretical uncertainties with forecast errors from a next-generation measurement combining DESI and a DETF4 type survey, the uncertainties from theory are of the same order of magnitude as, or less than, the predicted errors. However, in a more futuristic measurement, such as that from a combination of SKA2 and LSST, this is not the case. If it is our objective to use $E_G(R)$ as a tool for testing gravity in more futuristic scenarios, it will be essential to take these effects into account. 

In the case of the projection length $P$, one approach would be to compare the measured $E_G(R)$ with a prediction computed using our expression with the appropriate $P$ value. For the scale-dependent bias this method would be less practical, as the parameters of the bias models must be known. Thus, accurate modelling of and accounting for $b(k)$ is more likely the solution in this case. In the more futuristic scenario of using $E_G(R)$ not as a consistency check, but as the basis of a Bayesian analysis, this could be accomplished by parameterising $b(k)$ (via a functional form or simply by binning in $k$), and marginalising over the corresponding bias parameters. 

The case of theoretical uncertainty from scale-dependent bias is of particular interest, given that the original objective of $E_G$ was to circumvent degeneracy with the galaxy bias. We note that, as mentioned previously, the definition of $E_G(R)$ assumes that the measured $\beta(z)=f(z)/b(z)$ is scale-independent, which implies that the factor of the galaxy bias which enters via $\beta(z)$ is scale-independent by definition. The result is that any scale-dependent behaviour in the bias will fail to cancel. One potential way to address this would be to measure instead $\beta(z,k) = f(z,k) / b(z,k)$. If possible, this could lead to much less sensitivity of $E_G(R)$ to the galaxy bias. Additionally, we note that the effects of introducing scale-dependent bias are made more problematic by the real-space nature of the observational $E_G(R)$ than they would be in the case of the original Fourier-space statistic, in which case truncation at a particular value of $k$ could eliminate all effects. In light of this, it may be of interest to move towards a Fourier-space observable, as was originally proposed. This idea was explored for the case in which galaxy-galaxy lensing is replaced by CMB lensing in \cite{Pullen2014, Pullen2015}. 

In addition to the general relativistic expression for $E_G(R)$, we have also introduced an expression for $E_G(R)$ in the quasistatic limit of modified gravity. This has allowed us to compare deviations from the GR value of $E_G(R)$ with forecast errors. We find that, for a measurement from SKA2+LSST, we expect $E_G(R)$ to provide $1\sigma$ constraints on modified gravity parameters $\bar{\mu}_0$ and $\Sigma_0$ at a level of roughly $20\%$. This level of constraint is significantly looser than that expected from a combination of cosmic shear tomography and measurements of the growth rate and of baryon acoustic oscillations combined directly from a DETF4 type survey. Bear in mind, however, that we have focused on one redshift bin; by cleverly dividing the data into a few redshift bins, it should be possible to increase the statistical power of this method. Furthermore, one primary advantage of $E_G(R)$ is its use of galaxy-galaxy lensing, and the resulting relative insensitivity to cosmic shear systematics. Hence, even a reduced constraint from these different observational techniques will provide a useful cross-check. We also recall that our use of an effective redshift for the lens galaxy population may inflate forecast errors (see Appendix \ref{section:variance}), and hence this represents an upper bound on the expected error. Finally, we have found that although $E_G(R)$ uses both galaxy-galaxy lensing and galaxy-clustering measurements, $E_G(R)$ is primarily sensitive to $\Sigma_0$, with more limited sensitivity to changes in $\bar{\mu}_0$. 

$E_G(R)$ as an observable has the potential to provide a powerful test gravity, complementary to constraints using standard likelihood methods. However, in order to benefit from the measurement of $E_G(R)$ from more futuristic surveys, we must understand and model the theoretical uncertainties of the statistic. Our work here has been a step towards this goal, in providing full theoretical expressions for $E_G(R)$ in GR and in the quasistatic limit for modified gravity, and in studying the theoretical uncertainties which arise in comparison with forecast errors for futuristic surveys.

\section*{Acknowledgements}
We would like to thank Lance Miller, David Alonso, Thibaut Louis, Elisa Chisari, and Chris Blake for helpful discussions. We thank also the anonymous referee for his or her comments and suggestions. We additionally thank the authors of the publicly available code CAMB, which was used in this work. CDL is funded by the Rhodes Trust and the Natural Sciences and Engineering Research Council of Canada. PGF is supported by STFC, the Beecroft Trust and the Oxford Martin School. CH acknowledges support from the European Research Council under grant numbers 647112 and 204185.

\appendix
\section{Forecast errors on $E_G(R)$}
\label{section:variance}
\noindent
Here we present the calculation of the forecast errors on $E_G(R)$, binned in R. For each bin $R_i$, the variance of $E_G^i$ is given by:
\begin{equation}
\sigma^2(E_G^i) = (E_G^i)^2 \left( \left(\frac{\sigma(\Upsilon_{gm}^i)}{\Upsilon_{gm}^i}\right)^2 + \left(\frac{\sigma(\Upsilon_{gg}^i)}{\Upsilon_{gg}^i}\right)^2 + \left(\frac{\sigma(\beta)}{\beta}\right)^2 \right).
\label{sepvar}
\end{equation}
We therefore calculate the error of $\Upsilon_{gm}^i$, $\Upsilon_{gg}^i$ and $\beta$ in turn. All practical error calculations assume the fiducial choices of $R_0$, $W_s$, and $b$, while for $P$ we take the more realistic choice of $300$ Mpc/h.

\subsection{Forecast errors on $\Upsilon_{gm}$}
\label{subsection:upgmvar}
\noindent
In order to calculate the forecast error of $\Upsilon_{gm}(R)$ in bin $R_i$ (which we will denote $\sigma(\Upsilon_{gm}^i)$) we follow the method of \cite{Jeong2009}, where a similar calculation is made for the tangential shear. Consider $\Upsilon_{gm}(R)$ as written in the form:
\begin{align}
\Upsilon_{gm}(R)&=\Delta \Sigma_{gm}(R)-\left(\frac{R_0}{R}\right)^2\Delta \Sigma_{gm}(R_0).
\label{ygm_again}
\end{align}
We first calculate the variance of $\Delta \Sigma_{gm}$, then extend this result to $\Upsilon_{gm}$. $\Delta \Sigma_{gm}$ is given by:
\begin{align}
\Delta \Sigma_{gm}(\vec{R}) &= \frac{1}{\bar{w}}\int_0^{\chi_{\infty}} d \chi_s W_s(\chi_s) \overline{\Sigma_c^{-1} }(\bar{\chi}_l, \chi_s) \gamma_t^g(\bar{\chi}_l, \chi_s, \vec{R})
\label{DeltaSigma}
\end{align}
where we have not yet azimuthally averaged and $\vec{R}=(R\cos\phi, R\sin\phi)$. Note that $\gamma_t^g$ here is the tangential shear {\it about a lens galaxy}. 
 
The variance of $\Delta \Sigma_{gm}$ is given by:
\begin{equation}
\langle \Delta \Sigma_{gm}(R) \Delta \Sigma_{gm}(R') \rangle - \langle \Delta \Sigma_{gm}(R) \rangle \langle \Delta \Sigma_{gm}(R') \rangle.
\label{varDeltaSig}
\end{equation} 
We will first compute $\langle \Delta \Sigma_{gm}(\vec{R}) \Delta \Sigma_{gm}(\vec{R'}) \rangle - \langle \Delta \Sigma_{gm}(\vec{R}) \rangle \langle \Delta \Sigma_{gm}(\vec{R'}) \rangle$, then azimuthally average. 

Consider the first term in equation \ref{varDeltaSig}. Following equation B1 of \cite{Jeong2009}, we can write:
\begin{align}
\langle &\Delta \Sigma_{gm}(\vec{R}) \Delta \Sigma_{gm}(\vec{R'}) \rangle = \frac{1}{\bar{w}^2}\int_0^{\chi_\infty} d \chi_s W_s(\chi_s) \int_0^{\chi_\infty} d\chi_s' W_s(\chi_s') \frac{1}{N_L^2} \sum_{i,j}^{N_L} \int d^2 \hat{n} \int d^2 \hat{n}' \nonumber \\ & \times \langle \delta(\hat{n}-\hat{n}^i)  \delta(\hat{n'}-\hat{n'}^j) \overline{\Sigma_c^{-1}}(\bar{\chi}_l, \chi_s) \overline{\Sigma_c^{-1}}(\bar{\chi}_l, \chi_s') \gamma_t^g(\hat{n}, \bar{\chi}_l, \chi_s, \vec{R}) \gamma_t^g(\hat{n'}, \bar{\chi}_l, \chi_s', \vec{R'}) \rangle
\label{varterm1_1}
\end{align}
where $\hat{n}$ represents the direction on the sky at which the lens galaxy is situated, and $N_L$ is the total number of lens galaxies 
The sum above separates into two cases: $i=j$ and $i \ne j$. This results in two terms, representing the case where two shears are measured relative to the same galaxy and the case where they are measured relative to different galaxies:
\begin{align}
\langle \Delta &\Sigma_{gm}(\vec{R}) \Delta \Sigma_{gm}(\vec{R'}) \rangle = \frac{1}{\bar{w}^2}\int_0^{\chi_\infty} d \chi_s W_s(\chi_s) \int_0^{\chi_\infty} d\chi_s' W_s(\chi_s')  \frac{1}{N_L^2} \sum_{i,j}^{N_L} \int d^2 \hat{n} \int d^2 \hat{n}' \nonumber \\ & \times \Bigg[ \delta(\hat{n}-\hat{n'}) \langle n_L(\hat{n}, \bar{\chi}_l)  \overline{\Sigma_c^{-1}}(\bar{\chi}_l, \chi_s) \overline{\Sigma_c^{-1}}(\bar{\chi}_l, \chi_s')\gamma_t(\hat{n}, \chi_s, \vec{R}) \gamma_t(\hat{n'}, \chi_s', \vec{R'}) \rangle \nonumber \\ &+ \langle n_L(\hat{n}, \bar{\chi}_l) n_L(\hat{n}', \bar{\chi}_l)\overline{\Sigma_c^{-1}}(\bar{\chi}_l, \chi_s) \overline{\Sigma_c^{-1}}(\bar{\chi}_l, \chi_s') \gamma_t(\hat{n}, \chi_s, \vec{R}) \gamma_t(\hat{n'},  \chi_s', \vec{R'}) \rangle \Bigg].
\label{varterm1_2}
\end{align}
$n_L(\hat{n})$ is the surface density of lens galaxies at direction $\hat{n}$. Note that here the dependence of the tangential shear on $\bar{\chi}_l$ has vanished because $\gamma_t$ is not the tangential shear about a galaxy, but just the tangential shear. We expand: $n_L(\hat{n})=\bar{n}_L(1+\delta_g(\hat{n},\bar{\chi}_l))$, and use $N_L=\bar{n}_L 4 \pi f_{sky}$ to get:
\begin{align}
\langle &\Delta \Sigma_{gm}(\vec{R}) \Delta \Sigma_{gm}(\vec{R'}) \rangle = \frac{1}{4\pi f_{sky} N_L \bar{w}^2} \int_0^{\chi_\infty} d \chi_s W_s(\chi_s) \int_0^{\chi_\infty} d\chi_s' W_s(\chi_s') \nonumber \\ & \times \int d^2 \hat{n} \overline{\Sigma_c^{-1}}(\bar{\chi}_l, \chi_s) \overline{\Sigma_c^{-1}}(\bar{\chi}_l, \chi_s')\langle   \gamma_t(\hat{n}, \chi_s, \vec{R}) \gamma_t(\hat{n}, \chi_s', \vec{R'}) \rangle \nonumber \\ &+ \frac{1}{f_{sky}^2 16 \pi^2\bar{w}^2}\int_0^{\chi_{\infty}} d \chi_s W_s(\chi_s) \int_0^{\chi_{\infty}} d\chi_s' W_s(\chi_s') \int d^2 \hat{n} \int d^2 \hat{n}'\overline{\Sigma_c^{-1}}(\bar{\chi}_l, \chi_s) \overline{\Sigma_c^{-1}}(\bar{\chi}_l, \chi_s') \nonumber \\ &\times \Big[\langle \gamma_t(\hat{n}, \chi_s, \vec{R}) \gamma_t(\hat{n'},\chi_s', \vec{R'}) \rangle  +\langle \delta_g(\hat{n}, \bar{\chi}_l) \delta_g(\hat{n}', \bar{\chi}_l) \gamma_t(\hat{n}, \chi_s, \vec{R}) \gamma_t(\hat{n'}, \chi_s', \vec{R'}) \rangle \Big]
\label{varterm1_3}
\end{align}
As in \cite{Jeong2009}, we assume $\langle \delta_g \gamma_t \gamma_t \rangle=0$.

First we consider the one galaxy term (the first term of equation \ref{varterm1_3}). We Fourier expand the quantities inside angle brackets to get:
\begin{align}
& \frac{1}{\bar{w}^2}\int_0^{\chi_\infty} d \chi_s W_s(\chi_s) \int_0^{\chi_\infty} d\chi_s' W_s(\chi_s')\overline{\Sigma_c^{-1}}(\bar{\chi}_l, \chi_s) \overline{\Sigma_c^{-1}}(\bar{\chi}_l, \chi_s') \nonumber \\ &\times \Bigg[\frac{1}{N_L} \int \frac{d^2 \vec{l}}{(2\pi)^2} P_{\kappa\kappa}(l, \chi_s, \chi_s') \cos[2(\phi-\varphi)] \cos[2(\phi'-\varphi)]e^{i\vec{l}(\vec{R}-\vec{R'})} +\frac{\sigma_\gamma^2\delta(\vec{R}-\vec{R}')}{N_L n_s} \Bigg]
\label{onehalo_1}
\end{align}
where $\vec{l}=(l\cos\varphi, l\sin\varphi)$, $P_{\kappa\kappa}$ is the 2D power spectrum of the convergence , and $n_s$ is the average source galaxy surface density.

Now consider the two-galaxy contribution. The first part of the two-galaxy term is zero, being just two averages of tangential shear over the sky. The second part can be expanded using Wick's theorem. We get:
\begin{align}
\langle \Delta &\Sigma_{gm}(\vec{R})\rangle \langle \Delta \Sigma_{gm}(\vec{R'}) \rangle +\frac{1}{4\pi f_{sky}\bar{w}^2}\int_0^{\chi_\infty} d \chi_s W_s(\chi_s)  \int_0^{\chi_\infty} d\chi_s' W_s(\chi_s') \int d^2 \hat{n} \int d^2 \hat{n}'  \nonumber \\ & \times \overline{\Sigma_c^{-1}}(\bar{\chi}_l, \chi_s)   \overline{\Sigma_c^{-1}}(\bar{\chi}_l, \chi_s') \int \frac{d^2 \vec{l}}{(2\pi)^2}  \cos[2(\phi-\varphi)]\cos[2(\phi'-\varphi)]  e^{i\vec{l}(\vec{R}-\vec{R'})} \nonumber \\ &\times \Bigg[P_{g\kappa}(l, \bar{\chi}_l, \chi_s')P_{g\kappa}(l, \bar{\chi}_l, \chi_s)  + P_{gg}(l, \bar{\chi}_l)\Bigg[ P_{\kappa\kappa}(l, \chi_s, \chi'_S) + \frac{\sigma_\gamma^2}{n_s} \Bigg]\Bigg]
\label{2halo_1}
\end{align}
We see that the first part of this term cancels with $ \langle \Delta \Sigma_{gm}(\vec{R})\rangle \langle \Delta \Sigma_{gm}(\vec{R'}) \rangle$ in equation \ref{varDeltaSig}.

Adding together equations \ref{2halo_1} and \ref{onehalo_1} gives $\langle \Delta \Sigma_{gm}(\vec{R}) \Delta \Sigma_{gm}(\vec{R}') \rangle$. We then subtract $ \langle \Delta \Sigma(\vec{R})\rangle \langle \Delta \Sigma(\vec{R'}) \rangle$ to get the variance of $\Delta \Sigma_{gm}(\vec{R})$, and azimuthally average to get:
\begin{align}
\langle &\Delta \Sigma_{gm}(R) \Delta \Sigma_{gm}(R') \rangle - \langle \Delta \Sigma_{gm}(R) \rangle \langle \Delta \Sigma_{gm}(R') \rangle = \nonumber \\ &\frac{1}{4\pi f_{sky}\bar{w}^2}\int_0^{\chi_\infty} d \chi_s W_s(\chi_s) \int_0^{\chi_\infty} d\chi_s' W_s(\chi_s')   \overline{\Sigma_c^{-1}}(\bar{\chi}_l, \chi_s) \overline{\Sigma_c^{-1}}(\bar{\chi}_l, \chi_s')  \int \frac{l dl}{2\pi} J_2(lR) J_2(lR') \nonumber \\ &\times \Bigg[P_{g\kappa}(l, \bar{\chi}_l,  \chi_s')P_{g\kappa}(l, \bar{\chi}_l, \chi_s) + \left(P_{gg}(l, \bar{\chi}_l,)+\frac{1}{\bar{n}_L}\right)\Bigg[ P_{\kappa\kappa}(l, \chi_s, \chi_s') + \frac{\sigma_\gamma^2}{n_s} \Bigg]\Bigg]
\label{finalDeltaSigma}
\end{align}
where we have used $N_L=\bar{n}_L4\pi f_{sky}$ to incorporate the one-halo term. We then simply average in $R$ bins:
\begin{align}
\sigma^2 \left(\Delta \Sigma_{gm} ^i\right) &= \frac{1}{4\pi \bar{w}^2f_{sky}(R^{i+1}-R^i)(R^{i+1}-R^i) }\int_0^{\chi_\infty} d \chi_s W_s(\chi_s)    \int_0^{\chi_\infty} d\chi_s' W_s(\chi_s') \nonumber \\ & \overline{\Sigma_c^{-1}}(\bar{\chi}_l, \chi_s) \overline{\Sigma_c^{-1}}(\bar{\chi}_l, \chi_s')  \int_{R_i}^{R_{i+1}} dR \int_{R_i}^{R_{i+1}} dR' \int \frac{l dl}{2\pi} J_2(lR) J_2(lR')  \nonumber \\ &\times  \Bigg[P_{g\kappa}(l, \bar{\chi}_l, \chi_s')P_{g\kappa}(l, \bar{\chi}_l, \chi_s) + \left(P_{gg}(l, \bar{\chi}_l)+\frac{1}{\bar{n}_L}\right)\Bigg[ P_{\kappa\kappa}(l, \chi_s, \chi_s') + \frac{\sigma_\gamma^2}{n_s} \Bigg]\Bigg].
\label{finalDeltaSigma_ave}
\end{align}

We finally approximate the error on the second term in equation \ref{ygm_again} by computing the variance of $\Delta \Sigma_{gm}(R)$ in a small (extent $R_0/5$) bin around $R_0$, and then averaging the full term in $R$ bins. The total squared $1\sigma$ error of $\Upsilon_{gm}(R)$ in bin $R_i$ is the sum of this result and equation \ref{finalDeltaSigma_ave}.

We note here that, as in the main text, we have assumed an effective redshift $\bar{z}_l$ for the lens population, corresponding to an effective comoving distance $\bar{\chi}_l$. However, in computing forecast errors, we must account for the fact that some source galaxies may be at lower redshift than $\bar{z}_l$. In a real observational scenario, these galaxies would still be lensed by low redshift lens galaxies, and hence would still contribute to the signal. However, in this case, they cannot be considered. We compute the fraction of galaxies from the source galaxy distribution that are at higher redshift than $\bar{z}_l$, and we multiply $n_s$ by this fraction before computing errors. Thus, error bars are likely somewhat inflated and should be treated as an upper bound.

\subsection{Forecast errors on $\Upsilon_{gg}$}
\noindent
$\Upsilon_{gg}(R)$ as given in equation \ref{ygg} consists of three terms, all related to the projected correlation function $w_{gg}(R)$. A measurement of $w_{gg}(R)$ is made by first measuring the redshift-space correlation function $\xi^s_{gg}(R,\Delta)$. When calculating expected errors, we assume that as long as our projection length is long, it will be equivalent to use the real space correlation function, $\xi_{gg}(R, \Delta)$. We have that the variance of this correlation function is given by:
\begin{align}
\sigma^2&(\xi_{gg}(R_1, \Delta_1, R_2, \Delta_2)) = \frac{1}{\pi^2 V} \int_0^{\infty} dk_{||} \cos(k_{||} \Delta_1) \cos(k_{||} \Delta_2) \nonumber \\ &\times \int_0^{\infty} dk_{\perp} k_{\perp} J_0(k_\perp R_1) J_0(k_\perp R_2) \left(P_{gg}(k_{||}, k_\perp) + \frac{1}{n_g}\right)^2,
\label{varxigg}
\end{align} 
where $n_g$ is the volume density of galaxies and $V$ is the volume of the survey. We project $\Delta_1$ and $\Delta_2$ over $\{-P, \, P\}$, and average over bin $R_i$, to find:
\begin{align}
\sigma^2&(w_{gg}^i) = \frac{4}{(\Delta R_i)^2 \pi^2 V} \int_{R^i}^{R^{i+1}} dR_1 \int_{R^i}^{R^{i+1}} dR_2  \int_0^{\infty}dk_{||} \frac{\sin^2(k_{||} P)}{k_{||}^2} \nonumber \\ &\times \int_0^{\infty} dk_{\perp} k_{\perp} J_0(k_\perp R_1) J_0(k_\perp R_2) \left(P_{gg}(k_{||}, k_\perp) + \frac{1}{n_g}\right)^2.
\label{wggRvar}
\end{align}
It is numerically difficult to compute this quantity directly in the above form. This is as a result of the $\frac{1}{n_g^2}$ term: performing the integrals in $k_\perp$ over the oscillatory Bessel functions is unfeasible without the damping effect of the smoothed power spectrum at high $k_\perp$. Fortunately, we can expand the bracketed term and evaluate the $\frac{1}{n_g^2}$ portion analytically. We do so, and find:
\begin{align}
\sigma^2&(w_{gg}^i) = \frac{4}{(\Delta R_i)^2 \pi^2 V} \int_{R^i}^{R^{i+1}} dR_1 \int_{R^i}^{R^{i+1}} dR_2 \int_0^{\infty}dk_{||} \frac{\sin^2(k_{||} P)}{k_{||}^2}  \int_0^{\infty} dk_{\perp} k_{\perp} J_0(k_\perp R_1) \nonumber \\ &\times J_0(k_\perp R_2) \left( P_{gg}(k_{||}, k_\perp)^2 + 2P_{gg}(k_{||}, k_\perp)/n_g \right)+ \frac{2 |P| \ln\left(\frac{R^{i+1}}{R^i}\right)}{\pi V \Delta R_i n_g^2}
\label{wggRvar_RD}
\end{align}
where $\Delta R_i$ is the extent of the given $R$ bin. The variance of the middle term of equation \ref{ygg} is trivially the above times $\rho_c^2$. 

To find the squared error associated with the first term of equation \ref{ygg}, we use the same strategy, and find:
\begin{align}
\sigma^2 &= \frac{16}{(\Delta R_i)^2 \pi^2 V} \int_{R^i}^{R^{i+1}} dR_1 \int_{R^i}^{R^{i+1}} dR_2 \frac{1}{R_1^2 R_2^2} \times \int_{R_0}^{R_1} dR'_1\int_{R_0}^{R_2} dR'_2 \int_0^{\infty}dk_{||} \frac{\sin^2(k_{||} P)}{k_{||}^2} \nonumber \\ &\times \int_0^{\infty} dk_{\perp} k_{\perp} J_0(k_\perp R'_1) J_0(k_\perp R'_2) \left( P_{gg}(k_{||}, k_\perp)^2 + \frac{2P_{gg}(k_{||}, k_\perp)}{n_g} \right) \nonumber \\ &+ \frac{4 |P|}{(\Delta R_i)^2 \pi n_g^2 V}\left(\frac{1}{R^i}-\frac{1}{R^{i+1}}\right)\left(R^{i+1}-R^i + \frac{R_0^2}{2}\left(\frac{1}{R^{i+1}}-\frac{1}{R^i}\right) \right)
\label{wggintvar}
\end{align}

We estimate the error on the third term of equation \ref{ygg}, $\frac{R_0^2}{R^2}w_{gg}(R_0)$, by approximating the variance of $w_{gg}(R_0)$ as the variance in a small bin around $w_{gg}(R_0)$ (with extent $R_0/5$).

\subsection{Forecast errors on $\beta$}
\noindent
In computing the error on $\beta$, we employ the formula developed in \cite{Bianchi2012}:
\begin{equation}
\sigma(\beta) = \beta \frac{C b^{0.7}}{V^{0.5}}e^{\frac{n_0}{b^2n_g}}
\label{sigmab}
\end{equation}
where $C=4.5\times10^2 \rm{h}^{-1.5} \rm{Mpc}^{1.5}$ and $n_0=1.7\times10^{4} \rm{h}^3 \rm{Mpc} ^{-3}$. $b$ is the constant galaxy bias.

\section{$\Upsilon_{gm}$ without the assumption of an effective redshift}
\label{section:nozeff}
\noindent
In the main text above, we derive $\Upsilon_{gm}(R)$ under the assumption of an effective redshift for the lens galaxies. Here, we give the expression for $\Upsilon_{gm}(R)$ with this assumption removed.  

We introduce a function to characterise the lens galaxy redshift distribution, $W_l(\chi)$, normalised over the integration range. Equation \ref{deltasigthe1} becomes: 
\begin{align}
\Delta\Sigma_{gm}(R)&=\frac{1}{\bar{w}} \int_0^{\chi_H} d\chi_s W_s(\chi_s)\int_0^{\chi_H} d\chi_l W_l(\chi_l) \overline{\Sigma_{c}^{-1}}(\chi_l, \chi_s) \gamma_t(R, \chi_l, \chi_s)
\label{deltasigapp}
\end{align}
with $\bar{w}$ becoming:
\begin{align}
\bar{w}&= \int_0^{\chi_H} d \chi_s W_s(\chi_s) \int_0^{\chi_H} d\chi_l W_l(\chi_l) \left(\overline{\Sigma_{c}^{-1}}(\chi_l, \chi_s) \right)^2.
\label{wbarapp}
\end{align}
We now extend equation \ref{kappa_guzik} in a similar way as in the main text to obtain an expression for the convergence in direction ${\hat \theta_2}$ and at $\chi_l$ given a galaxy in direction ${\hat \theta_1}$ and at $\chi_s$, in terms of two comoving distances and the angle $\theta$ between directions ${\hat \theta_1}$ and ${\hat \theta_2}$. We find:
\begin{align}
\kappa_g(\theta, \chi_l, \chi_s) &=\frac{3}{2}\left(\frac{H_0}{c}\right)^2\Omega_M(z=0) \int_0^{\chi_H} d\chi_l W_l(\chi_l) \int_0^{\chi_s}d\chi \frac{g(\chi,\chi_s)}{a(\chi)} [\xi(\chi_l,\chi, \theta)].
\label{kappat1t2app}
\end{align}
In order to convert this to a function of $R$, we write $\theta = \frac{R}{\chi_l}$, such that the value of $\theta$ depends on $\chi_l$ at each point in the lens distribution. Then propagating through to the expression for $\Upsilon_{gm}(R)$, we find:
\begin{align}
\label{ygmapp}
&\Upsilon_{gm}(R)= \frac{\rho_c \Omega_M(z=0)}{\bar{w}}\int_0^{\chi_H} d\chi_l W_l(\chi_l) \int d\Delta \frac{(4\pi G)^2 \chi_l (\chi_l+\Delta)}{c^4a(\chi_l) a(\chi_l+\Delta)} \int_{\chi_l+\Delta}^{\chi_H} d\chi_s \overline{W}_s(\chi_s) \nonumber \\ & \times\frac{(\chi_s-\chi_l-\Delta)(\chi_s-\chi_l)}{\chi_s^2}    \Bigg[\frac{2}{R^2}\int_{0}^{R}R'dR' \xi_{gm}\left(\chi_l, \chi_s, \frac{R'}{\chi_l}, b \right) \nonumber \\
& -\xi_{gm}\left(\chi_l, \chi_s, \frac{R}{\chi_l}, b \right)+\left(\frac{R_0}{R}\right)^2\xi_{gm}\left(\chi_l, \chi_s, \frac{R_0}{\chi_l}, b \right)\Bigg]
\end{align}


\providecommand{\href}[2]{#2}\begingroup\raggedright\endgroup

\end{document}